\documentclass[pra,onecolumn,preprint,11pt,showpacs,citeautoscript,footinbib,eqsecnum,superscriptaddress]{revtex4-2}
\usepackage[T1]{fontenc}
\usepackage[utf8]{inputenc}
\usepackage{lipsum}
\usepackage{amsmath,bm}
\usepackage{amssymb}
\usepackage{bbm}
\usepackage{braket}
\usepackage{graphicx}
\usepackage{xcolor}
\usepackage{pifont}
\usepackage[mathscr]{euscript}
\usepackage[shortlabels]{enumitem}
\usepackage[justification=justified]{subcaption}
\usepackage{float}
\usepackage{dsfont}
\usepackage{comment}
\usepackage{subcaption}
\usepackage{slashed}

\usepackage[papersize={8.5in,11in}]{geometry}
\usepackage[colorlinks=true]{hyperref}  
\hypersetup{
    unicode=false,          
    pdftoolbar=true,        
    pdfmenubar=true,        
    pdffitwindow=false,     
    pdfstartview={FitH},    
    pdftitle={Flstar},    
    pdfauthor={Christos, Sachdev, Luo},     
    pdfsubject={},   
    pdfcreator={},   
    pdfproducer={}, 
    pdfkeywords={} {} {}, 
    pdfnewwindow=true,      
    colorlinks=true,       
    linkcolor=blue, 
    citecolor=blue,        
    filecolor=magenta,      
    urlcolor=blue           
} 
\renewcommand{\vec}[1]{\boldsymbol{#1}}

\newcommand{\vi}{{\boldsymbol{i}}}
\newcommand{\vj}{{\boldsymbol{j}}}

\newcommand{\figref}[1]{Fig.\,\ref{#1}}

\newcommand{\eqnref}[1]{Eq.\,\eqref{#1}}

\renewcommand{\Im}{\mathop{\mathrm{Im}}}

\begin{document}
\graphicspath{{figures/}}

\title{Vortex structure in a $d$-wave superconductor\\ obtained by a confinement transition from the pseudogap metal}

\begin{abstract} 
We compute the structure of flux $h/(2e)$ vortices in a $d$-wave superconductor which emerges from a higher temperature pseudogap metal. Such a transition is described by a continuum theory of the Higgs condensation of 2 flavors of charge $e$ bosons which are fundamentals of an emergent SU(2) gauge field.
Period-2 charge order is present near the vortex center. Upon coupling the electrons to the superconducting and charge order parameters, we find that the electronic local density of states does not have a zero-bias peak, in contrast to BCS theory. But there are sub-gap peaks at positive and negative bias, and these exhibit anti-phase periodic spatial modulations, similar to observations in scanning tunneling microscopy experiments in the underdoped cuprates (K. Matsuba {\it et al.}, J. Phys. Soc. Jpn. {\bf 76}, 063704 (2007)).  
\end{abstract}

\author{Jia-Xin Zhang}
\affiliation{Institute for Advanced Study, Tsinghua University, Beijing 100084, China}
\affiliation{Department of Physics, Harvard University, Cambridge MA 02138, USA}

\author{Subir Sachdev}
\affiliation{Department of Physics, Harvard University, Cambridge MA 02138, USA}

\maketitle
\newpage
\linespread{1.05}
\tableofcontents

\section{Introduction}
\label{sec:intro}
It is now well established that the superconducting state in the hole-doped cuprates has $d$-wave pairing \cite{KirtleyRMP}. Moreover, the 4 nodal fermionic quasiparticle excitations observed by photoemission \cite{ShenRMP} match the Bogoliubov quasiparticles expected after imposing $d$-wave pairing on a `large' Fermi surface enclosing the Luttinger hole volume of $1+p$, where $p$ is hole-doping density away from the insulator.
However, discrepancies appear upon a closer look at the microstructure of the $d$-wave superconductor provided by its Abrikosov vortex: the pioneering scanning tunneling microscopy (STM) observations of Hoffman {\it et al.} \cite{science.1066974} were not in agreement with the Bogoliubov-de Gennes theory of a $d$-wave superconductor with an underlying large 
Fermi surface. A complete understanding of their and subsequent observations \cite{Fischer05,Nishida.Matsuba.2007,Nishida.Yoshizawa.2013,Tamegai.Machida.2016,Hamidian19}, and their connection to the normal state electronic spectrum at low doping, remain central open problems in the physics of the cuprates. 

Using the Bogoliubov-de Gennes theory, Wang and MacDonald \cite{WM95} predicted a large zero-bias peak  in the electronic local density of states (LDOS) at the core of each vortex. Such a peak was not seen in numerous STM experiments \cite{science.1066974,Fischer05,Nishida.Matsuba.2007,Nishida.Yoshizawa.2013,Tamegai.Machida.2016,Hamidian19}, but has finally been observed \cite{Renner21} in heavily overdoped Bi$_{2}$Sr$_{2}$CaCu$_{2}$O$_{8 +\delta}$; this observation is consistent with large Fermi surface present in the overdoped cuprates \cite{ShenRMP}. 

The focus of the present paper is on the underdoped cuprates, where the normal state above the superconducting critical temperature, $T_c$, is the `pseudogap metal'. This pseudogap metal does not have a large Fermi surface, and instead the photoemission spectrum is characterized by `Fermi arcs' \cite{ShenRMP} (we will discuss a specific model for the pseudogap metal below). We will study the consequences of this electronic spectrum for the vortex below $T_c$.
The STM investigations of the vortex structure \cite{science.1066974,Fischer05,Nishida.Matsuba.2007,Nishida.Yoshizawa.2013,Tamegai.Machida.2016,Hamidian19} in the underdoped cuprates have revealed a number of remarkable features, not consistent the Bogoliubov-de Gennes theory of $d$-wave superconductivity in a large Fermi surface Fermi liquid:
\begin{itemize}
\item There is no Wang-MacDonald zero-bias peak.
\item The zero-bias peak is replaced by sub-gap peaks at $\pm$6-9 meV in the LDOS \cite{Fischer05,Nishida.Matsuba.2007}.
\item The LDOS at the sub-gap peaks exhibits periodic spatial modulations in a `halo' around the vortex core. The modulations were initially observed to have a period of 4 lattice spacings \cite{science.1066974,Fischer05,Nishida.Matsuba.2007,Nishida.Yoshizawa.2013,Tamegai.Machida.2016}, but a recent experiment \cite{Hamidian19} has seen an additional period 8 modulation. Period 8 modulations have also been observed in superconducting states without vortices  \cite{Hirschfeld20,Fujita20,Seamus21}, and  have been interpreted by the co-existence of pair density waves and a uniform superconductor.  
\item There is an `anti-phase' relation \cite{Nishida.Matsuba.2007} between the LDOS modulations with positive and negative bias.
\end{itemize}

Many theories for the vortex structure in the underdoped regime \cite{Ting01,Demler02,Ogata03,Ting03,BBBSS1,Schmid_2010,Agterberg15,Kivelson18,Pepin20,Hirschfeld20,TKLee23,Ghosal02,Ghosal24,Datta2023} commonly involve modeling with effective actions for multiple competing or intertwined orders, including antiferromagnetism, $d$-wave superconductivity, charge density wave, and pair density wave orders. Often implicit in these approaches is the assumption that the pseudogap state above $T_c$ can be understood in terms of a theory of fluctuations of one or more of these orders \cite{KivelsonRMP,Lee2014,Lee2020}. But the specific theoretical computations of the vortex structure do not account for these order parameter fluctuations---hence, the normal state in these computations has a large Fermi surface.

We shall proceed here in the opposite direction, and regard the pseudogap metal as a quantum phase of matter in its own right, and as the `parent' of the ordered phases that emerge at low $T$.
We model the pseudogap metal as a `fractionalized Fermi liquid' (FL*) with hole pocket Fermi surfaces enclosing a volume $p$. Such a Fermi surface does not enclose the Luttinger volume, and then Oshikawa's argument \cite{MO00} for the Luttinger volume in a Fermi liquid implies \cite{FLS2} that a non-Luttinger volume Fermi surface must be accompanied by a spin liquid with fractionalized excitations.  We take this background spin liquid to be a SU(2) gauge theory of fermionic spinons moving in a background $\pi$-flux \cite{pnas.2302}. 
Then the low temperature $d$-wave superconducting, antiferromagnetic, charge density wave or pair density wave phases emerge upon transitions which confine the fractionalized excitations of the spin liquid. (We note that there are dual theories of this spin liquid which are expressed in terms of bosonic spinons \cite{BCS}, and which lead to related theories of the pseudogap metal \cite{SSROPP19}, but we find the fermionic spinon description appropriate for the purposes of the present paper.) Thus our approach captures {\it both} the normal state pseudogap, and the structure of the vortex in the $d$-wave superconductor that emerges at low temperature.

The ancilla method \cite{Ancilla20} provides a powerful tool to describe the FL* pseudogap metal and its confinement transitions.
Although it is possible to formulate our analysis without any reference to the ancilla method  \cite{ChatterjeeSS16}, we shall employ it as a convenient and foolproof method to connect to a lattice scale Hamiltonian, and to account for all anomalies in fractionalized phases. 
The ancilla method has been used in earlier work to describe\\
({\it i\/}) the electronic photoemission spectrum of the 
FL* pseudogap metal, with the `Fermi arcs' appearing as the front-sides of the hole pockets \cite{Mascot22}, \\
({\it ii\/}) the onset of a uniform $d$-wave superconductor with 4 nodal quasiparticles with anisotropic velocities from FL* \cite{ChatterjeeSS16,Christos24},\\
({\it iii\/}) and the quantum oscillations in the charge-ordered state at low $T$ and high fields \cite{BCS}.\\
Here, we shall use the ancilla method to describe the structure of the vortex that emerges upon the confinement transition from FL* to the $d$-wave superconductor.

This confinement is described by the condensation of a Higgs boson $B_\vi \equiv (B_{1\vi}, B_{2\vi})$ which is a fundamental of the gauge SU(2), and carries charge $e$ under the electromagnetic U(1); here $\vi$ denotes a square lattice site, and $B_{a\vi}$ ($a=1,2$ is a SU(2) gauge index) are complex scalars. Then gauge-invariant bilinears of $B_\vi$ determine the structure of the confining phase where $B_\vi$ is condensed. The independent on-site and nearest-neighbor bilinears are \cite{pnas.2302}:
\begin{align}\label{eq:order_lattice}
&\mbox{site charge density:~}\left\langle c_{\vi \alpha}^\dagger c_{\vi \alpha}^{\vphantom\dagger} \right\rangle \sim \rho_{\vi} = B^\dagger_\vi B_\vi^{\vphantom\dagger} \nonumber \\
&\mbox{bond density:~} \left\langle c_{\vi \alpha}^\dagger c_{\vj \alpha}^{\vphantom\dagger} + c_{\vj \alpha}^\dagger c_{\vi \alpha}^{\vphantom\dagger} \right\rangle~\sim Q_{\vi \vj} = Q_{\vj\vi} = \mbox{Im} \left(B^\dagger_\vi e_{\vi \vj}^{\vphantom\dagger} U_{\vi\vj}^{\vphantom\dagger} B_\vj \right) \nonumber \\
&\mbox{bond current:~} i\left\langle c_{\vi \alpha}^\dagger c_{\vj \alpha}^{\vphantom\dagger} - c_{\vj \alpha}^\dagger c_{\vi \alpha}^{\vphantom\dagger} \right\rangle \sim J_{\vi \vj} = - J_{\vj\vi} =  \mbox{Re} \left( B^\dagger_\vi e_{\vi \vj}^{\vphantom\dagger} U_{\vi \vj}^{\vphantom\dagger} B_\vj^{\vphantom\dagger} \right) \nonumber \\
&\mbox{Pairing:~} \left\langle \varepsilon_{\alpha\beta} c_{i \alpha} c_{j \beta} \right\rangle \sim \Delta_{\vi \vj} = \Delta_{\vj \vi} = \varepsilon_{ab} B_{a\vi} e_{\vi \vj} U_{\vi \vj} B_{b\vj}\,. 
\end{align}
Here $\vi,\vj$ are nearest-neighbors on the square lattice, $U_{\vi\vj}$ is the SU(2) lattice gauge field, $\varepsilon$ is the unit antisymmetric tensor, and we have specified the interpretation of the bilinears in terms of the underlying electron operator $c_{\vi\alpha}$ ($\alpha = \uparrow, \downarrow$ are SU(2) spin indices). The fixed field $e_{\vi\vj} = - e_{\vj \vi}$ specifies the $\pi$-flux of the spin liquid, and our choice is $e_{\vi, \vi + \hat{x}} = 1$, $e_{\vi, \vi + \hat{y}} = (-1)^x$.

A uniform $d$-wave superconductor has $\Delta_{\vi, \vi+\hat{x}} = -\Delta_{\vi, \vi+\hat{y}} $ but independent of $\vi$, $\rho_{\vi}$ and $Q_{\vi\vj}$ spatially uniform, and $J_{\vi\vj}=0$. `Charge order' can appear from spatial modulations in $\rho_\vi$, $Q_{\vi\vj}$, or $\Delta_{\vi\vj}$, and modulations in $\Delta_{\vi\vj}$ can be identified as pair density waves. But note that the configurations of these order parameters are all tied to the $B_\vi$. Thus we can regard $B_\vi$ as simultaneously a `square root' of $d$-wave superconductivity, charge density wave, and pair density wave orders, and the different orientations of $B_{\vi}$ describes the intertwining of these orders.

Ref.~\cite{pnas.2302} presented an effective lattice action for the $B_{\vi}$ and $U_{\vi\vj}$, along with its couplings to the fermionic spinons and the electronic quasiparticles. It would be of interest to study vortex lattice saddle points of this action in the presence of an applied magnetic field (similar to analyses in Ref.~\cite{TKLee23} for the Bogoliubov-de Gennes theory augmented by competing orders). However such a complete analysis would be numerically very demanding, and here we will limit ourselves to a simplified continuum limit for which analytic results are possible \cite{pnas.2302,CSSL24}.
The condensation of $B_{\vi}$ in this limit can lead to charge order only with period of two lattice spacings. Longer period charge orders are possible in the theory of Ref.~\cite{pnas.2302} (see also Ref.~\cite{BCS}), and we expect the main features of our results will generalize to more realistic situations. In this context, it is worth noting that for charge order periods longer than 2, the condensation of the $B_{\vi}$ is at wavevectors $\pm {\bm Q}_{\rm cdw}/2$, where ${\bm Q}_{\rm cdw}$ is the charge-ordering wavevector. In the presence of co-existing superconductivity, the orientation of $B_{\vi}$ can also induce `pair density wave' order at ${\bm Q}_{\rm cdw}/2$, and we propose this as an origin of the observed period 8 modulations \cite{Hamidian19,Hirschfeld20,Fujita20,Seamus21}.

We also note the relationship between our model for underdoped vortex structure and the analysis of Nagaosa, Lee, and Wen \cite{LeeWenRMP} (see also Refs.~\cite{SSvortex,NagaosaLee92,LeeWen01,BBBSS05,BBS06,BS07}).
These authors also considered the condensation of complex scalars with quantum numbers similar to $B_{\vi}$, but in the context of a different `staggered flux' spin liquid with a U(1) gauge field (which is unstable to a trivial U(1)-monopole \cite{Alicea08,Song:2018ial}). In their case, the condensation of $B_{\vi}$ leads to $d$-wave superconductivity or current order, but no charge order.

Our continuum model for the $B_\vi$ is described in Section~\ref{sec:continuum}. It is expressed in terms of 2 SU(2) fundamental complex fields $B_{as}$, where $s = \pm$ is a `valley' index, and $a$ is the previously specified SU(2) gauge index. For bosons, valleys are determined by the minimum of the dispersion, and for the gauge choice of $e_{\vi\vj}$ above, the valleys are chosen at wavevectors ${\bm Q}_+=\frac{\pi}{2}(1,1)$ and ${\bm Q}_-=\frac{\pi}{2}(1,-1)$.
We obtain analytic results for the vortex with flux $h/(2e)$ in the continuum theory and show that the vicinity of the vortex core has charge order (for suitable choices of the quartic self-interactions of the $B_{as}$). There are some similarities to this vortex-induced charge order to earlier analyses \cite{BBBSS05,BBS06,BS07} which used a dual model for the quantum fluctuations of the vortices.

Section~\ref{sec:fermions} turns to a computation of the electronic spectrum in the background of the vortex obtained in Section~\ref{sec:continuum}. We use a simplified continuum model in which the wavevector of the period-2 charge order is exactly equal to the separation between the nodal points of the superconductor. 
We find that the charge order has significant effects on the LDOS, and compare our results to STM data.

\section{Continuum model for Higgs fields $B_{\pm}$}
\label{sec:continuum}
\begin{figure}[t]
	\centering
    \includegraphics[width=0.6\linewidth]{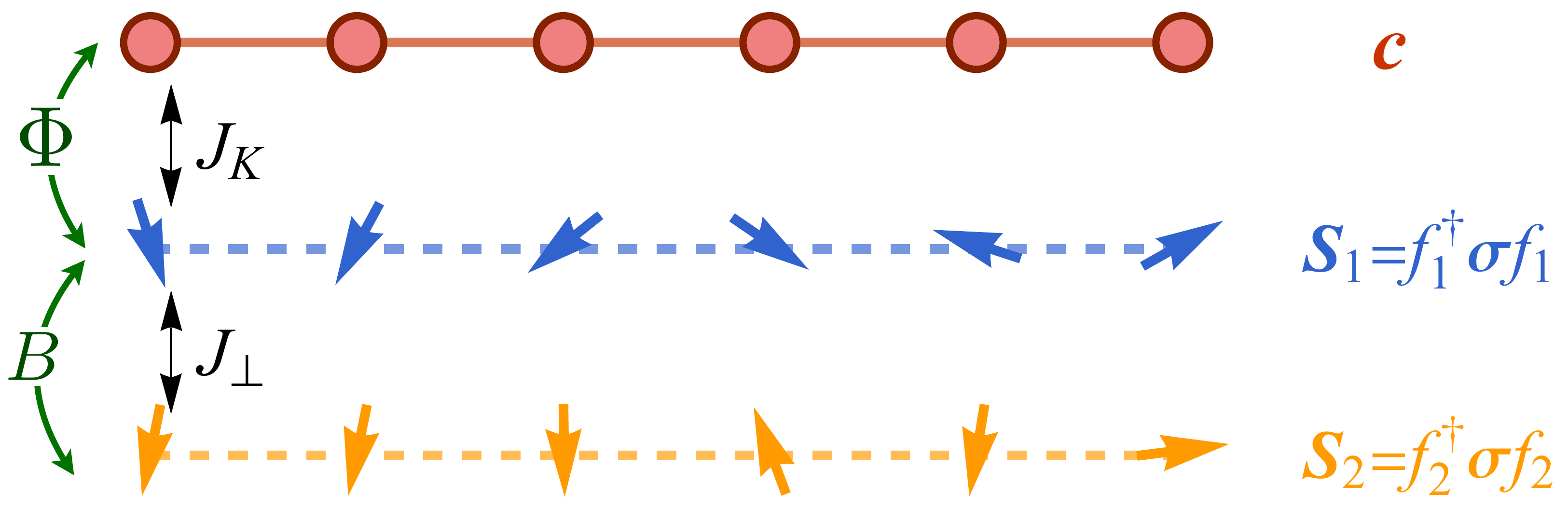}
	\caption{Schematic illustration of the ancilla model. Red circles denote the actual electrons, while blue arrows in the second layer and orange arrows in the third layer represent ancilla spins $\vec{S}_1$ and $\vec{S}_2$, respectively, which can be further fractionalized into spinons $f_1$ and $f_2$. Black arrows between distinct layers indicate Kondo coupling $J_K$ and rung-exchange $J_\perp$, respectively. The $B$ and $\Phi$ are Higgs bosons which couple the layers, as in (\ref{eq:H_ancilla}). } 
	\label{fig_ancilla}
\end{figure}

Here, we utilize the ancilla approach introduced in Ref. \cite{pnas.2302} to describe a single-band Hubbard model of electrons $c$ with on-site repulsion $U$. This on-site repulsion $U$ can be captured by coupling the $c$ electrons to an insulated bilayer square lattice antiferromagnet composed of spin-1/2 moments $\boldsymbol{S}_1$ and $\boldsymbol{S}_2$, as shown in \figref{fig_ancilla}. When the rung-exchange $J_{\perp}$ in the bilayer antiferromagnet is much larger than the Kondo interaction $J_K$ between the $c$ electrons and $\boldsymbol{S}_1$, the $\boldsymbol{S}_1$ and $\boldsymbol{S}_2$ spins form a trivial rung-singlet state. In this scenario, the $c$ electrons are largely decoupled from the ancilla spins and form a large Fermi surface with Luttinger volume $1+p$, corresponding to the Fermi liquid phase in the overdoped regime. Conversely, when $J_{\perp} \ll J_K$, the ancilla spins $\boldsymbol{S}_1$ are expected to be Kondo screened by the $c$ electrons, forming a Fermi pocket with a non-Luttinger volume $p$. Meanwhile, the $\boldsymbol{S}_2$ spins form a $\pi$-flux spin liquid. This latter case corresponds to the $\mathrm{FL}^*$ phase, which is the focus of this work.

The corresponding Hamiltonian of the ancilla model is given by:
\begin{eqnarray}\label{eq:H_ancilla}
\begin{aligned}
    H =&\sum_{\boldsymbol{i}, \boldsymbol{j}}\left[t_{\boldsymbol{i j}}^c c_{\boldsymbol{i}}^{\dagger} c_{\boldsymbol{j}}+t_{\boldsymbol{i j}}^f f_{1, \boldsymbol{i},\sigma}^{\dagger} f_{1, \boldsymbol{j},\sigma}\right] +i J \sum_{\langle\boldsymbol{i}, \boldsymbol{j}\rangle} f_{2, \boldsymbol{i}, \sigma}^{\dagger} e_{\boldsymbol{i}\boldsymbol{j}} f_{2, \boldsymbol{j}, \sigma}\\
    &+ \sum_i \left[\Phi c_{\boldsymbol{i}, \sigma}^{\dagger} f_{1, \boldsymbol{i}, \sigma}+i B_{1 \boldsymbol{i}} f_{2, \boldsymbol{i}, \sigma}^{\dagger} f_{1, \boldsymbol{i}, \sigma}-i B_{2 \boldsymbol{i}} \epsilon_{\alpha \beta} f_{2, \boldsymbol{i}, \alpha} f_{1, \boldsymbol{i}, \beta}+\text { H.c }\right],
    \end{aligned}
\end{eqnarray}
where $f_{1}$ and $f_{2}$ are the fermionic partons of the ancilla spins $\boldsymbol{S}_{1} = f_{1}^\dagger \boldsymbol{\sigma}f_{1}$ and $\boldsymbol{S}_{2} = f_{2}^\dagger \boldsymbol{\sigma}f_{2}$ (with $\boldsymbol{\sigma}$ as the Pauli matrix), respectively. Here, $t_{\boldsymbol{ij}}^c$ and $t_{\boldsymbol{ij}}^f$ represent the hopping terms for physical $c$ electrons and $f_{1}$, while $e_{\boldsymbol{ij}}$ denotes the $\pi$-flux hopping for $f_{2}$. Importantly, the hybridization field $\Phi$ and the chargon field $B$ with a unit electromagnetic charge are introduced to decouple the Kondo interaction $J_{K}$ and the rung-exchange $J_{\perp}$, respectively. The different phases of $H$ are:
\begin{itemize}
    \item The FL* pseudogap metal phase is obtained when $\Phi \neq 0$ and $B=0$: then the $c$ and $f_1$ fermions hybridize to form the small hole pockets of size $p$, while the $f_2$ fermions realize the `spectator' $\pi$-flux spin liquid.
    \item The Higgs/confinement transition to the superconductor is described by the condensation of the Higgs field $B$ in a background with $\Phi \neq 0$. We study here the nature of this condensation in the presence of a $h/(2e)$ vortex in the superconducting phase. The choice between the order parameters in Eq.~(\ref{eq:order_lattice}) is made by the orientation of the complex doublet $(B_{1\vi}, B_{2\vi})$.
    \item The large Fermi surface Fermi liquid phase (FL) is obtained when $\Phi=0$ and $B=0$, and the $f_{1,2}$ fermions confine to form a rung-singlet phase. We do not study this FL phase or the FL-FL* transition \cite{Ancilla20,Ancilla20b} in the present paper.
\end{itemize}

Following the analysis in Refs.~\cite{pnas.2302,CSSL24}, the low-lying modes of the $B$ field at the continuum limit can be regarded as a 4-component bosonic field. Two of these components are the gauge $SU(2)$ components, as shown in \eqnref{eq:H_ancilla}, while the other two correspond to two degenerate dispersion minima (two distinct valleys). Hence, the effective free energy of the chargon $B$ in the continuum limit can be obtained by integrating out all the other fields in \eqnref{eq:H_ancilla}, resulting in:
\begin{eqnarray}\label{totfree}
    F&=&\int d^2 \boldsymbol{r}\left|\left(\nabla+i \boldsymbol{A}^e+i \boldsymbol{a}^z \sigma_z\right) B_{s}\right|^2+m B_{s}^{\dagger} B_{s}+u\left(B_s^{\dagger} B_{s}\right)^2\\
    &\;&+v_1\left(\rho_{(\pi, 0)}^2+\rho_{(0, \pi)}^2\right)+v_2 D^2+v_3|\Delta|^2 +\frac{1}{g}\left(\nabla \times \boldsymbol{A}^e\right)^2+\frac{1}{\lambda}\left(\nabla \times \boldsymbol{a}^z\right)^2,\notag,
\end{eqnarray}
where the relevant conventional order parameters, consisting of bilinear terms of the $B$ fields, are defined as:
\begin{eqnarray}\label{eq:order_con}
    d\text{-wave SC}: &\;&\Delta=\epsilon_{a b} B_{a+} B_{b-} \notag\\
    x\text{-CDW}: &\;& \rho_{(\pi, 0)}=B_{a+}^{\dagger} B_{a+}-B_{a-}^{\dagger} B_{a-}\label{order}\\
    y\text{-CDW}: &\;& \rho_{(0, \pi)}=B_{a+}^{\dagger} B_{a-}+B_{a-}^{\dagger} B_{a+} \notag\\
    d\text{-density wave}: &\;& D=i\left(B_{a+}^{\dagger} B_{a-}-B_{a-}^{\dagger} B_{a+}\right).\notag
\end{eqnarray}
Here, $\Delta$ represents the $d$-wave superconducting (SC) order, $\rho_{(\pi,0)}$ and $\rho_{(0, \pi)}$ represent the period-2 charge density wave (CDW) along the $x$ and $y$ directions, respectively, while $D$ represents the $d$-density wave order, which breaks time-reversal symmetry and is characterized by a circulating current pattern. In these equations, $s=\pm$ and $a=1,2$ represent the valley index and the gauge $SU(2)$ index of the $B$ field, respectively. It is important to note that the CDW states above are primarily in the bond densities, $Q_{\vi\vj}$, in Eq.~(\ref{eq:order_lattice}) \cite{pnas.2302,CSSL24}; so when co-existing with superconductivity this will lead to pair density wave order represented by modulations in 
$\Delta_{\vi\vj}$.

As the mass $m$ becomes negative, confinement occurs via the condensation of the chargon field $B$, leading to distinct possible conventional orders as defined in \eqnref{eq:order_con}. The specific phase of the system is determined by the parameters $v_1$, $v_2$, and $v_3$ in \eqnref{totfree}. For our subsequent discussions, we choose these parameters so that the bulk system is in the $d$-wave superconducting phase. We hypothesize a solution by adopting an ansatz for $B_\pm$ in the SC phase:
\begin{equation}\label{eq:uniSC}
    B_{+}=b\left(\begin{array}{c}
\cos \theta \\
\sin \theta
\end{array}\right), \;\;\;\;
B_{-}=b\left(\begin{array}{c}
-\sin \theta \\
\cos \theta
\end{array}\right)
\end{equation}
where $b$ and $\theta \in [0,2\pi)$ are both real numbers. This leads to $\Delta=\epsilon_{a b} B_{a+} B_{b-}=b^2$ and $\rho_{(\pi, 0)}=\rho_{(0, \pi)}=D=0$, indicating a homogeneous superconducting phase without any coexisting order.

Next, we will explore the superconducting vortex states in the presence of an external magnetic field $\boldsymbol{A}^e$, and we will show that the vortex physics is intimately related to the internal gauge field $\boldsymbol{a}^z$, with coupling constants introduced by $\lambda$ in \eqnref{totfree}. It is critical to note that while, in principle, bosonic chargons $B_\pm$ carry fundamental SU(2) gauge charges (i.e., $i \boldsymbol{a}^j \sigma_j$), only the $z$-component is considered in \eqnref{totfree} for simplicity. This simplification is justified by the fact that the three components of gauge fluctuation are completely frozen as the Higgs field $B_\pm$ condenses in the superconducting phases.

\begin{table}
    \centering
    \begin{tabular}{c|c|c|c|c}
         & $B_{1+}$ & $B_{2+}$ & $B_{1-}$ & $B_{2-}$\\
         \hline
        $\vec{A}^e$ & $2\pi $ & $2\pi $ & $2\pi $ & $2\pi $\\
        \hline
        total & $2\pi $ & $2\pi $ & $2\pi $ & $2\pi $ \\
    \end{tabular}
    \caption{With the flat configuration of $\vec{a}^z$, the $\boldsymbol{A}^e$-flux and total flux experienced by $B$ fields far from the vortex core.}
    \label{tab:my_label}
\end{table}

\subsection{Flat configuration}
As $\lambda \rightarrow 0$, the cost of finite $\vec{a}^z$ flux tends to infinity, so the flat gauge structure predominates, signifying $\vec{a}^z=0$. In this scenario, the components of $B_{\pm}$ can perceive the flux of $\boldsymbol{A}^e$, as summarized in Table \ref{tab:my_label}. Consequently, compared with \eqnref{eq:uniSC}, the ansatz for the chargon $B$ is modified to:

\begin{equation}\label{anaB}
    B_{+}=b\left(\begin{array}{c}\cos \theta \\ \sin \theta\end{array}\right) f(r) e^{i \phi}, \;\;\;\; B_{-}=b\left(\begin{array}{c}-\sin \theta \\ \cos \theta\end{array}\right) f(r) e^{i  \phi}
\end{equation}
where $f(r) \in [0,1]$ represents the amplitude modulation along the radial direction, and $\phi$ is the coordinate angle in the two-dimensional real space, encircling the pole $r=0$ with $2\pi$ winding. Note that, in general, phase winding can be any integer multiple of $2\pi$ to satisfy the single-valued condition for the charge $e$ boson $B_{\pm}$. However, we only consider the lowest energy vortex in this context, corresponding to a $2\pi$ phase winding. Here, $b$ and $\theta$, which are inherited from \eqnref{eq:uniSC}, are independent of spatial coordinates.

\begin{figure}[t]
	\centering
    \includegraphics[width=0.9\linewidth]{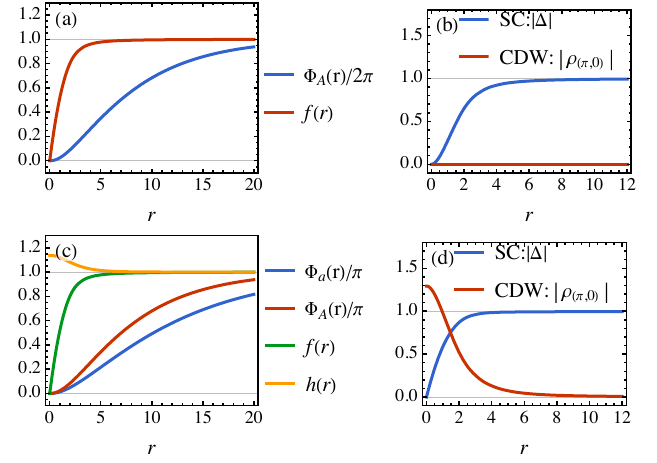}
	\caption{(a)-(b) With the parameters: $m=-1$, $u=2$, $v_3=-1$, $g=10u$. Spatial dependence of $\Phi_A(r)$ and $f(r)$ with the chargon $B$ ansatz in \eqnref{anaB} is shown in (a), and the corresponding magnitude of order parameter $\Delta$ and $\rho_{(\pi,0)}$ in \eqnref{order1} in (b). (c)-(d) With the parameters: $m=-1$, $u=2$, $v_1=0.5$, $v_3=-1$, $g=10u$, $\lambda = g/9$. Spatial dependence of $\Phi_a(r)$, $\Phi_A(r)$, $f(r)$, and $h(r)$ with the chargon $B$ ansatz in \eqnref{Bpm2} is shown in (c), and the corresponding magnitude of order parameter $\Delta$ and $\rho_{(\pi,0)}$ in \eqnref{order2} in (d).} 
	\label{fig_trivial}
\end{figure}

Then, as shown in \figref{fig_trivial}(a), one can determine the spatial dependence of $f(r)$ and $\Phi_A(r)$ by optimizing the free energy \eqnref{totfree} after replacing the $B$ field with the ansatz \eqnref{anaB}; see Appendix~\ref{App01}. These results reveal that $f(r)$ is suppressed near the vortex core to prevent the divergence of kinetic energy. Additionally, the electromagnetic flux $\Phi_A(r) = \oint A^e(r) \cdot d \vec{l}$ asymptotically approaches $2\pi$ ($=h c / e \equiv 2\phi_0$ if the full units are restored) at distances far from the vortex core. Therefore, the order parameters, as defined in \eqnref{order}, are represented by:
\begin{eqnarray}
    d\text{-wave SC}: &\;&\Delta=b^2 f^2(r) e^{i 2 \phi} \notag\\
    x\text{-CDW}: &\;& \rho_{(\pi, 0)}=0\label{order1}\\
    y\text{-CDW}: &\;& \rho_{(0, \pi)}=0 \notag\\
    d\text{-density wave}: &\;& D=0,\notag
\end{eqnarray}
This configuration implies the formation of a $4\pi$ superconducting vortex, attributable to the phase $2\phi$, while all the other intertwined orders completely vanish. The spatial variation of the magnitudes of these order parameters is presented in \figref{fig_trivial}(b), indicating that the superconducting order parameters diminish in proximity to the vortex core. Consequently, the trivial flat SU(2) gauge structure is capable of yielding $4\pi$ superconducting vortices exclusively, without the coexistence of any other order.

\subsection{Vison-like configuration}\label{vlc}
On the other hand, as $\lambda$ becomes large, contrary to the flat configuration, the existence of a vison-like configuration is allowed, characterized by a $\pi$ flux structure in the gauge field $\boldsymbol{a}^z$:
\begin{equation}\label{piflux}
    \Phi_a(r)=\oint \boldsymbol{a}^z(r) \cdot d \vec l\xrightarrow{r \rightarrow \infty} \pi.
\end{equation} 
Due to the first term in \eqnref{totfree}, $B_{as}$ with the opposite index of $a$ perceive opposite fluxes of $\boldsymbol{a}^z$. Thus, the resulting total gauge structure perceived by $B_{as}$ is summarized in Table \ref{tab:total}, suggesting that $B_{1s}$ experiences a total $2\pi$ flux, leading to the $2\pi$ phase winding, while $B_{2s}$ does not need to have any phase winding. In this context, the single-valued condition can still be well satisfied. Similarly, the $B_{as}$ can be expressed as:
 \begin{equation}\label{Bpm2}
    B_{+}=b\left(\begin{array}{c}\cos \theta f(r) e^{i  \phi} \\ \sin \theta h(r)\end{array}\right) , \;\;\;\; B_{-}=b\left(\begin{array}{c}-\sin \theta  f(r) e^{i  \phi}\\ \cos \theta h(r)\end{array}\right).
\end{equation}
Based on the ansatz of \eqnref{Bpm2}, the results of the spatial dependence for $\Phi_a(r)$, $\Phi_A(r)$, $f(r)$, and $h(r)$ are depicted in \figref{fig_trivial}(c), illustrating that $f(r)$ is suppressed near the vortex core, while $h(r)$ remains approximately constant; see Appendix~\ref{App02}. The fluxes of both $\Phi_A$ and $\Phi_a$ converge to $\pi$ ($=h c / 2 e \equiv \phi_0$ if the full units are restored) at distances far from the vortex core, satisfying the minimal flux quantization condition in superconductors. Then, the order parameters defined in \eqnref{order} can be expressed as:
\begin{eqnarray}
    d\text{-wave SC}: &\;&\Delta=b^2 f(r) h(r) \boldsymbol{e}^{i \phi} \notag\\
    x\text{-CDW}: &\;& \rho_{(\pi, 0)}=\left(f^2(r)-h^2(r)\right) b^2 \cos 2 \theta \label{order2}\\
    y\text{-CDW}: &\;& \rho_{(0, \pi)}=\left(f^2(r)-h^2(r)\right)b^2 \sin 2 \theta \notag\\
    d\text{-density wave}: &\;& D=0,\notag
\end{eqnarray}
This configuration implies the formation of a $2\pi$ superconducting vortex and the potential emergence of stripe-like CDW. The spatial variation in the magnitudes of these order parameters is shown in \figref{fig_trivial}(d), indicating that the superconducting order parameters are suppressed near the vortex core, while CDW $\rho_{(\pi, 0)}$ and $\rho_{(0, \pi)}$ become prominent within the vortex core.

As a result, in scenarios where $\lambda$ is not negligible, the existence of a vison-like structure in the gauge field $\vec{a}^z$ facilitates the formation of a $2\pi$ superconducting vortex, in which the core induces a local charge density wave. Notice that a excitation carrying $\pi$ flux `trapped' in the vortex is also necessary under the framework of other parton constructions \cite{LeeWen01, Muthukumar2002, Song-Weng2023} to obtain the correct flux quantization condition.

\begin{table}
    \centering
    \begin{tabular}{c|c|c|c|c}
         & $B_{1+}$ & $B_{2+}$ & $B_{1-}$ & $B_{2-}$\\
         \hline
        $\vec{A}^e$ & $\pi$ & $\pi$ & $\pi$ & $\pi$\\
        \hline
        $\vec{a}^z$ & $\pi$ & $-\pi$ & $\pi$ & $-\pi$ \\
        \hline
        total & $2\pi$ & $0$ & $2\pi$ & $0$ \\
    \end{tabular}
    \caption{With the vison-like configuration of $\vec{a}^z$, the $\boldsymbol{A}^e$-flux, $\boldsymbol{a}^z$-flux, and total flux perceived by $B$ fields far from the vortex core.}
    \label{tab:total}
\end{table}

\section{Electronic spectrum near the vortex}
\label{sec:fermions}

This section will further study the single-particle spectrum near the vortex based on the configuration of the $B$ field obtained in the last section at the continuum limit. The behavior of physical electrons in \eqnref{eq:H_ancilla} can be described effectively after integrating out ancilla spinons $f_1$ and $f_2$, and then working with an effective Hamiltonian of the electrons coupled to the physical, SU(2) gauge-invariant order parameters. In this respect, the analysis in the present section is similar to earlier competing order theories \cite{Ting01,Demler02,Ogata03,Ting03,BBBSS1,Schmid_2010,Agterberg15,Kivelson18,Pepin20,Hirschfeld20,TKLee23,Ghosal02,Ghosal24,Datta2023}. However, our order parameters descend from the theory for the fractionalized particle $B$ in Section~\ref{sec:continuum}, and the gauge-invariant order parameters in   
Eq.~(\ref{eq:order_lattice}). And as emphasized in Section~\ref{sec:intro}, the normal state of our theory is a pseudogap metal, not a large Fermi surface.

\subsection{Effective electronic Hamiltonian}

\begin{figure}[t]
	\centering
	\includegraphics[width=0.8\linewidth]{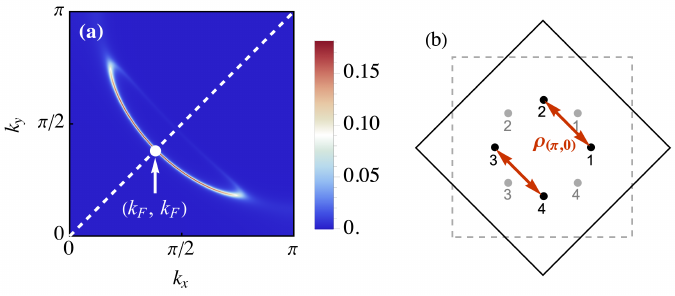}
	\caption{With parameters $t^c=200$, $(t^c)^\prime=-34$, $(t^c)^{\prime\prime}=-36$, $(t^c)^{\prime\prime\prime}=-7$, $t^f=-100$, $(t^f)^\prime=30$, $(t^f)^{\prime\prime}=10$, and doping density $\delta=0.2$. (a) Spectral function of \eqnref{spec} at zero frequency. (b) Labeling of the four Dirac points, shown in gray before and in black after rotation, as described in \eqnref{rotate}. Red arrows indicate the x-CDW coupling $\rho_{(\pi,0)}$ connecting different Dirac points.} 
	\label{spectrum}
\end{figure}

After integrating out the ancilla spinons, the resulting electronic Hamiltonian is given by:
\begin{equation}\label{Hc}
    H_c \equiv \sum_{k}H_0 c_{\vec{k},\sigma}^\dagger c_{\vec{k},\sigma}+\frac{\Phi^2}{\gamma^2 J} \sum_{\langle i, j\rangle}\left(\left(J_{\boldsymbol{ij}} + i Q_{\boldsymbol{ij}} \right) c_{\boldsymbol{i} \sigma}^{\dagger} c_{\boldsymbol{j}  \sigma}+\Delta_{\boldsymbol{i}\boldsymbol{j} } \epsilon_{\alpha \beta} c_{\boldsymbol{i} \alpha} c_{\boldsymbol{j} \beta}\right)+\text { H.c. },
\end{equation}
where $\gamma = -(\epsilon_{\vec{k}=0}^{f_1})^{-1}$ and $H_0$ is frequency-dependent with the expression as:
\begin{equation}
    H_0(\vec{k}, \omega)=\epsilon_{\vec{k}}^c+\frac{\Phi^2}{\omega-\epsilon_{\vec{k}}^{f_1}},
\end{equation}
where $\epsilon_{\boldsymbol{k}}^c$ and $\epsilon_{\boldsymbol{k}}^{f_1}$ are the bare dispersions for electrons $c$ and spinons $f_1$, respectively. These dispersions include the nearest neighbor terms ($t^c$, $t^f$), next nearest neighbor terms ($(t^c)'$, $(t^f)'$), next-next nearest neighbor terms ($(t^c)''$, $(t^f)''$), and next-next-next nearest neighbor terms ($(t^c)'''$) to fit the photo-emission data. Therefore, the corresponding spectral function
\begin{equation}\label{spec}
    A_c(\vec{k}, \omega)=-\frac{1}{\pi} \operatorname{Im} \frac{1}{\omega+i 0^{+}-H_{0}(\vec{k}, \omega+i 0^+)},
\end{equation}
is shown in \figref{spectrum} (a) at $\omega = 0$, demonstrating Fermi arc behaviors. Moreover, the interaction couplings $\Delta_{\boldsymbol{ij}}$, $J_{\boldsymbol{ij}}$, and $Q_{\boldsymbol{ij}}$ in \eqnref{Hc} are combinations of the $B$ field on the lattice [as defined in \eqnref{eq:order_lattice}], and their relation with continuum order parameters in \eqnref{eq:order_con} are given by:
\begin{eqnarray}\label{Deltaij}
     \begin{aligned}
  \Delta_{\boldsymbol{ij}} &=\eta_{\boldsymbol{ij}} 2(1+\sqrt{2}) \Delta \\   J_{\boldsymbol{ij}}&=-(1+\sqrt{2}) \rho_{(0,0)}+(-1)^{i_y}(2+\sqrt{2}) \rho_{(0, \pi)} \frac{1-\eta_{\boldsymbol{ij}}}{2}+(-1)^{i_x}(2+\sqrt{2}) \rho_{(\pi, 0)} \frac{1+\eta_{\boldsymbol{ij}}}{2}\\
        Q_{\boldsymbol{ij}}&=-(-1)^{i_x+i_y}(1+\sqrt{2})  D \eta_{\boldsymbol{ij}}   
     \end{aligned} 
    \end{eqnarray}
where $\eta_{\boldsymbol{i}, \boldsymbol{i} \pm \hat{x}} = -1$ and $\eta_{\boldsymbol{i}, \boldsymbol{i} \pm \hat{y}} = 1$. According to \eqnref{Deltaij}, it is explicitly shown that $\Delta_{\boldsymbol{ij}}$ represents the $d$-wave pairing potential, $J_{\boldsymbol{ij}}$ represents the period-2 charge order, and $Q_{\boldsymbol{ij}}$ corresponds to the current pattern, which is consistent with the symmetry analysis in Ref. \cite{pnas.2302}. Also, the distributions of the magnitudes of these order parameters near the vortex have been analyzed and obtained in \figref{fig_trivial} (d). Importantly, the electronic Hamiltonian of \eqnref{Hc} is explicitly gauge invariant, making it a suitable starting point for our subsequent discussion on the behavior of the electronic spectrum near vortices.

\subsection{Vortex states of Dirac fermions}
In the absence of the electromagnetic field, both $J_{\boldsymbol{ij}}$ and $Q_{\boldsymbol{ij}}$ vanish in a homogeneous $d$-wave superconducting phase. The Hamiltonian in \eqnref{Hc} in momentum space can then be expressed as:
\begin{equation}\label{HDelta}
    H_c = \sum_{\vec{k}} H_{0} c_{\vec{k}, \sigma}^{\dagger} c_{\vec{k}, \sigma}+\sum_{\vec{k}} \Delta_{\vec{k}} \epsilon_{\alpha \beta} c_{\vec{k}, \alpha} c_{-\vec{k}, \beta}+\text{H.c.}
\end{equation}
with $\Delta_{\vec{k}} = \Delta_0 \times\left(-\cos k_x + \cos k_y\right)$ and 
$\Delta_0 \equiv 4(1+\sqrt{2})J^{-1} \Phi^2 \gamma^2 \Delta$. 
Notably, the self-energy in $H_{0}(\omega)$, i.e., $\Sigma(k,\omega) = \Phi^2 /\left(\omega - \epsilon_{\vec{k}}^{f_1}\right)$, is small at the momentum $(k_F, k_F)$, 
leading to the approximation $H_{0} \simeq \epsilon_k^c$ near $(\pm k_F, \pm k_F)$. 
For simplicity, the momentum space will be rotated by $\pi/4$:
\begin{eqnarray}
    \left(\begin{array}{l}
        k_x \\
        k_y
        \end{array}\right) &\rightarrow& =\frac{1}{\sqrt{2}}\left(\begin{array}{cc}
        1 & -1 \\
        1 & 1
        \end{array}\right)\left(\begin{array}{l}
        k_x \\
        k_y
        \end{array}\right)\label{rotate},
\end{eqnarray}
resulting in the four Dirac points at the following momenta:
\begin{eqnarray}
    \vec{q}_1=\left(\sqrt{2} k_F, 0\right), \;\;\; \vec{q}_2=\left(0, \sqrt{2} k_F\right), \;\;\; \vec{q}_3=\left(-\sqrt{2}  k_F, 0\right), \;\;\; \vec{q}_4=\left(0,-\sqrt{2} k_F\right)
\end{eqnarray}
as depicted in \figref{spectrum}(b). Expanding near each Dirac point $\vec{q}_\alpha$ such that $\vec{k} = \vec{q}_\alpha + \vec{p}$, under the Nambu basis $\psi_\alpha(\vec{p}) = \left(c_{\alpha \uparrow}(\vec{p}) \;\; c_{\alpha \downarrow}^{\dagger}(-\vec{p})\right)^T$, the Hamiltonian in \eqnref{HDelta} can be rewritten as $H_c = \sum_\alpha \int d^2\vec{p} \, \psi_\alpha^{\dagger}(\vec{p}) h_\alpha(\vec{p}) \psi_\alpha(\vec{p})$, with $c_{\alpha \sigma}(\vec{p}) \equiv c_\sigma\left(\vec{q}_\alpha + \vec{p}\right)$, where $\alpha$ denotes the index of the Dirac points. For now, we focus on the node $\alpha = 1$ and postpone the discussion of the other nodes for later. In the presence of coupling to the external magnetic vector potential $\vec{A}^e$, the corresponding Hamiltonian matrix for this Dirac point with $\alpha = 1$ is given by:
\begin{equation}\label{h01}
    h_1=v_F p_x \sigma_z-v_F A_x^e \sigma_0+v_{\Delta} p_y \sigma_x
\end{equation}
where $v_{\Delta} = \Delta_0 \sqrt{2} k_F$ and $\sigma_x$, $\sigma_y$, $\sigma_z$ are Pauli matrices while $\sigma_0$ is the unit $2\times 2$ matrix. So far, the complicated `Fermi arc'-like electronic behaviors $H_0$ have been reduced to four free Dirac fermions in the $d$-wave SC phase. Next, based on this simplified description, we will continue the discussion of vortex physics in the presence of a vortex field.

Based on the analysis in Sec. \ref{vlc}, we established that in the presence of a magnetic field, around the $2\pi$ vortex, the $d$-density wave potential $Q_{\boldsymbol{ij}}$ in \eqnref{Deltaij} always vanishes, while a local charge order potential $J_{\boldsymbol{ij}}$ in \eqnref{Deltaij} is induced. First, we will temporarily ignore the local charge density potential, focusing on the problem of Dirac fermions perceiving a $\pi$-flux. In the following calculation, the suppression of the SC order parameter magnitude inside the vortex core will be ignored since the size of the vortex core in cuprates is so small compared to the size of the magnetic halo. 

Now, the Franz-Tešanović transformation $U$ acting on the electronic Hamiltonian \eqnref{HDelta} is applicable here \cite{Volovik96,Volovik97,Mel99,Mel2001, Tesanovic.Franz.2001, Tesanovic.Juricic.2008, Tesanovic.Franz.1999, Tesanovic.Vafek.2001, Tesanovic.Vafek.20019ri, Bartosch.Nikolic.2006, Sachdev.Nikolic.2006, Tesanovic.Melikyan.2007}:
\begin{eqnarray}
    U^{-1} H_c U &=&\sum_{\boldsymbol{k}}\left(\begin{array}{cc}
            e^{-i \Phi_A} \epsilon_{\boldsymbol{k}-\boldsymbol{A}^e}^c e^{i \Phi_A} & e^{-i \Phi_A} e^{i \phi / 2} \Delta_{\boldsymbol{k}} e^{i \phi / 2} e^{-i \Phi_B} \\
            e^{i \Phi_B} e^{-i \phi / 2} \Delta_{\boldsymbol{k}} e^{-i \phi / 2} e^{i \Phi_A} & -e^{-i \Phi_B} \epsilon_{-(\boldsymbol{k}+\boldsymbol{A}^e)}^c e^{i \Phi_B}
            \end{array}\right)\label{FTtrans}\\
            &=& \sum_{\boldsymbol{k}}\left(\begin{array}{cc}
                \epsilon_{\boldsymbol{k}-\boldsymbol{A}^e+\nabla \Phi_A}^c & \Delta_{\vec{k}+\vec{\mathcal{A}}} \\
                \Delta_{\vec{k}+\vec{\mathcal{A}}} & -\epsilon_{-\left(\vec{k}+\vec{A}^e-\nabla \Phi_B\right)}^c
                \end{array}\right)\label{sumk}
\end{eqnarray}
where 
\begin{equation}
    U=\left(\begin{array}{cc}
        e^{i \Phi_A} & 0 \\
        0 & e^{-i \Phi_B}
        \end{array}\right)
\end{equation}
and $\phi$ in \eqnref{FTtrans} represents the coordinate angle, arising from the $2\pi$ superconducting vortex. Here, we assume that $\Phi_A + \Phi_B = \phi$, and the new `gauge field' $\boldsymbol{\mathcal{A}} = \left(\nabla \Phi_A - \nabla \Phi_B\right) / 2$ in \eqnref{sumk} satisfies the relation $\oint d \boldsymbol{l} \cdot \boldsymbol{\mathcal{A}} = \pi$. Such effective gauge field $\boldsymbol{\mathcal{A}}$, tied to the phase gradient of the SC order parameter, corresponds to the $\pi$ flux at the vortex core, which enforces a statistical interaction between vortices and quasiparticles, causing a sign change in the quasiparticle’s wave function as it completes a loop around the vortex.


Following this transformation $U$, the Hamiltonian matrix of \eqnref{h01} can be expressed as:
\begin{equation}\label{h1FT}
h_1=v_F\left(p_x+\mathcal{A}_x\right) \sigma_z+v_F V_x \sigma_0+v_{\Delta}\left(p_y+\mathcal{A}_y\right) \sigma_x,
\end{equation}
where $\boldsymbol{V} \equiv \left(\nabla \Phi_A + \nabla \Phi_B\right) / 2 - \boldsymbol{A}^e$ denotes the supercurrent velocity. The terms with $\mathcal{A}_{x,y}$ account for the Aharonov-Bohm phase acquired by the quasiparticle around the vortex, while the $V_x$ term is a Doppler shift in the quasiparticle energy as a result of the background superfluid flow around the vortex \cite{Volovik96,Volovik97,Mel99,Mel2001}.
Both types of terms have been examined in earlier work \cite{Mel2001,Bartosch.Nikolic.2006,Sachdev.Nikolic.2006}: it was found that both induce a $1/r$ enhancement of the local density of states in the vortex core, with main difference being that the Doppler-shift contribution is anisotropic in space \cite{Mel2001}. In the interests of simplicity we will therefore not include the Doppler shift further here, and we do not expect this omission to lead to any qualitative change in our conclusions. It will be useful to confirm this with more detailed numerical work in the future.

Without the Doppler term, we can renormalize the length scales in Eq.~(\ref{h1FT}):
\begin{eqnarray}
    x \rightarrow v_F x, \;\;\;\;\;\;\;\; y \rightarrow v_{\Delta} y \label{scale}.
\end{eqnarray}
Following the unitary transformation, the rescaled Hamiltonian matrix of \eqnref{h1FT} can be expressed as:
\begin{equation}
    h_1=\left(p_x+\mathcal{A}_x\right) \sigma_x+\left(p_y+\mathcal{A}_y\right) \sigma_y.
\end{equation}
Switching to polar coordinates,
\begin{eqnarray}
    \begin{aligned}
        & p_x=\cos \phi\left(-i \frac{\partial}{\partial_r}\right)-\frac{\sin \phi}{r}\left(-i \frac{\partial}{\partial_\phi}\right) \\
        & p_y=\sin \phi\left(-i \frac{\partial}{\partial_r}\right)+\frac{\cos \phi}{r}\left(-i \frac{\partial}{\partial_\phi}\right)
        \end{aligned}
\end{eqnarray}
and
\begin{eqnarray}
    \begin{aligned}
        & \mathcal{A}_x=-\frac{\sin \phi}{2 r} \\
        & \mathcal{A}_y=\frac{\cos \phi}{2 r}
        \end{aligned}
\end{eqnarray}
the Bogoliubov-de Gennes (BdG) equation for $\alpha = 1$ with $q_1 = \left(k_F, 0\right)$ takes the form:
\begin{equation}\label{bdg}
    \left(\begin{array}{cc}
        0 & e^{-i \phi}\left[-i \frac{\partial}{\partial_r}-\frac{i}{r}\left(-i \frac{\partial}{\partial_\phi}+\frac{1}{2}\right)\right] \\
        e^{i \phi}\left[-i \frac{\partial}{\partial_r}+\frac{i}{r}\left(-i \frac{\partial}{\partial_\phi}+\frac{1}{2}\right)\right]
        \end{array}\right)\left(\begin{array}{c}
        u(r, \phi) \\
        v(r, \phi)
        \end{array}\right)=E\left(\begin{array}{c}
        u(r, \phi) \\
        v(r, \phi)
        \end{array}\right).
\end{equation}
By using the ansatz $u(r, \phi) = e^{i(l-1) \phi} u(r)$ and $v(r, \phi) = e^{i l \phi} v(r)$, one obtains the normalized wavefunction with the corresponding energy $E_{q, k} = q k$:
\begin{eqnarray}\label{uvcons}
\left(\begin{array}{l}
    u_{k, l, q, \alpha=1}(r,\phi) \\
    v_{k, l, q, \alpha=1}(r,\phi)
    \end{array}\right)=\sqrt{\frac{q k}{4 \pi}}\left(\begin{array}{c}
    e^{i(l-1) \phi} J_{s_l\left(l-\frac{1}{2}\right)}(k r) \\
    s_l i q e^{i l \phi} J_{s_l\left(l+\frac{1}{2}\right)}(k r)
    \end{array}\right).\label{wave1}
\end{eqnarray}
where $s_l = \operatorname{sgn}(l)$ with $s_{l=0} = -1$ and $J$ are Bessel functions of the first kind. Here, the states are labeled by three quantum numbers: radial wave vector $k > 0$, angular momentum $l \in \mathds{Z}$, and $q = \pm$, which characterize the particle and hole branches of excitations.

\subsection{Induced local charging order near the vortex}
Next, we will study how the induced local CDW potential $J_{\boldsymbol{ij}}$ influences the electronic spectrum near the vortex. Noting that $x$-CDW $\rho_{(\pi,0)}$ and $y$-CDW $\rho_{(0,\pi)}$ are degenerate, we will let $\rho_{(0,\pi)}=0$ in the following calculation for simplicity. Thus, from \eqnref{Deltaij}, the CDW potential in Hamiltonian \eqnref{Hc} is given by:
\begin{eqnarray}
\begin{aligned}
    &-\frac{\Phi^2}{\gamma^2 J} \sum_{\langle i, j\rangle}\left(J_{\boldsymbol{ij}} c_{\vec{i}, \sigma}^{\dagger} c_{\vec{j}, \sigma}+\text{H.c.}\right)\\
    &\approx-\frac{\Phi^2 \gamma^2 }{J} (2+\sqrt{2})\sum_\sigma \sum_{\alpha=1,3} \int d \vec{r} \rho_{(\pi, 0)}(\vec r)\left( c_{\alpha, \sigma}^{\dagger}(\boldsymbol{r}) c_{\alpha+1, \sigma}(\boldsymbol{r})+\text{H.c.} \right)\\
    &\approx-\frac{\Phi^2 \gamma^2 }{J} (2+\sqrt{2})\int d \boldsymbol{r} \rho_{(\pi, 0)}(\vec r)\int d k^{\prime} d k \sum_{l, l^\prime} \sum_{q, q^{\prime}}\sum_{\alpha=1,3}\left[u_{k^{\prime}, l^{\prime}, q^{\prime}, \alpha}^*(\vec r)+i v_{k^{\prime}, l^\prime, q^{\prime}, \alpha}^*(\vec r)\right]\\
    &\;\;\;\;\;\;\;\;\;\;\;\;\;\;\;\;\;\;\;\;\;\;\;\;\;\;\;\;\;\;\;\;\;\;\;\;\;\;\;\;\;\;\;\;\;\;\;\;\;\times\left[u_{k, l, q, \alpha+1}(\vec r)-i v_{k, l, q, \alpha+1}(\vec r)\right] \gamma_{k^{\prime}, l^\prime, q^{\prime}, \alpha}^{\dagger} \gamma_{k, l, q, \alpha+1}+\text{h.c.}\label{HCDW}
\end{aligned}
\end{eqnarray}
where $c_{\alpha,\sigma}$ is the physical electron near the Dirac point with the index $\alpha$, while $\gamma$ represents the Bogoliubov quasiparticles characterized by the quantum numbers $k, l, q, \alpha$, with the corresponding wave function of $(u, v)$ obtained in \eqnref{uvcons}. To ultimately obtain an analytic result, the following approximations are made here:
\begin{itemize}
    \item The momentum of Dirac points deviates slightly from $(\pm \pi/2, \pm \pi/2)$ as shown in \figref{spectrum}(b), but we will position it exactly at $(\pm \pi/2, \pm \pi/2)$ to let the $x$-CDW $\rho_{(\pi,0)}$ and $y$-CDW $\rho_{(0,\pi)}$ potentials connect different Dirac points perfectly.
    \item We assume that the size of the vortex halo is so large that the CDW potential can be approximately regarded as spatially independent inside the region we are concerned with, i.e., $\rho_{(\pi, 0)}(\boldsymbol{r}) \rightarrow \rho_{(\pi, 0)}$, allowing modes with different $k$ to decouple due to the orthogonality relation of Bessel functions.
    \item The rotational symmetry is broken when the coupling between different Dirac points leads to a coupling term between modes with $l$ and $l \pm 1$. We will neglect the off-diagonal term of $l$ in the following calculation. 
\end{itemize}

Calculations of more realistic situations without these approximations must be done numerically and depend on some microscopic details, i.e., the specific form of spatial decay of the CDW potential. However, we expect that there will be no significant qualitative modifications since the final analytic consequence we obtained mainly arises from the mixture between different flavors of Dirac fermions.

Based on these approximations, we can plug the wavefunction \eqnref{wave1} into \eqnref{HCDW}, then the electronic Hamiltonian \eqnref{Hc} can be expressed as:
\begin{equation}\label{Hcfinal}
    H_c=\int_0^{\infty} d k \sum_l \sum_{\alpha=1,3} \left(\Psi_{k, l}^{(\alpha)}\right)^{\dagger} V_{k, l}^{(\alpha)} \Psi_{k, l}^{(\alpha)}
\end{equation}
with the basis:
\begin{equation}
\Psi_{k, l}^{(\alpha)} = \left(\begin{array}{llll}
\gamma_{k, l, q=+, \alpha} & \gamma_{k, l, q=+, \alpha+1} &\gamma_{k, l, q=-, \alpha}& \gamma_{k, l, q=-, \alpha+1}
\end{array}\right)^T
\end{equation}
and the corresponding matrix
\begin{eqnarray}
    V_{k, l}^{(\alpha)}&=&=k \sigma_{30}-(\alpha-2)F s_l\left(\sigma_{31}+\sigma_{32}-\sigma_{11}-\sigma_{12}\right),
\end{eqnarray}
where $\sigma_{ij} = \sigma_i \otimes \tau_j$, with $\sigma$ and $\tau$ being Pauli matrices representing the Dirac point index and particle/hole branch with $q = \pm$, respectively. The parameter $F = -\Phi^2 \gamma^2(2 + \sqrt{2}) \rho_{(\pi, 0)}/4J$ is associated with the CDW potential.

Next, we will calculate the LDOS near the superconducting vortex in the presence of the local charging potential. According to \eqnref{Hcfinal}, Bogoliubov modes from different Dirac points have an overlap, so the LDOS $\rho(\omega, r)$ can be decomposed into a uniform part $\rho_{\text{uni}}(\omega, r)$ and a modulation part $\rho_{\mathrm{CDW}}(\omega, r)$:
\begin{eqnarray}\label{rhototal}
    \rho(\omega, r)=\rho_{\mathrm{uni}}(\omega, r)+\rho_{\mathrm{CDW}}(\omega, r) \times (-1)^r,
\end{eqnarray}
where the uniform part $\rho_{\text{uni}}(\omega, r)$ comes from the combination of the same Dirac fermions:
\begin{eqnarray}
    \rho_{\text {uni }}(\omega, r) \equiv \frac{1}{\pi} \sum_\sigma \sum_{\alpha=1}^4 \operatorname{Im}\left\langle c_{\alpha, \sigma}(r) c_{\alpha, \sigma}^{\dagger}(r)\right\rangle
\end{eqnarray}
and the modulation part $\rho_{\mathrm{CDW}}(\omega, r)$ with the spatial oscillation factor $(-1)^r$ comes from the combination of distinct Dirac fermions:
\begin{eqnarray}
    \rho_{\mathrm{CDW}}(\omega, r) \equiv \frac{1}{\pi} \sum_\sigma \sum_{\alpha=1,3} \operatorname{Im}\left[\left\langle c_{\alpha, \sigma}(r) c_{\alpha+1, \sigma}^{\dagger}(r)\right\rangle+ \left\langle c_{\alpha+1, \sigma}(r) c_{\alpha, \sigma}^{\dagger}(r)\right\rangle\right]
\end{eqnarray}

\begin{figure}[t]
	\centering
	\includegraphics[width=0.6\linewidth]{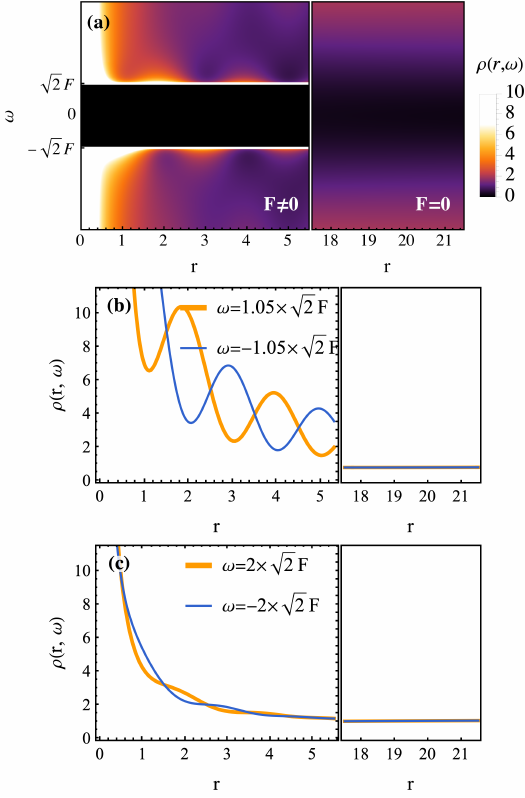}
	\caption{With the parameters $c=1/10$ and $F=0.3$, LDOS $\rho(r,\omega)$ near the vortex from \eqnref{LDOSfinal} is plotted in the left panel, and $\rho(r,\omega)_{\infty}$ far from the vortex from \eqnref{LDOSinf} is plotted in the right panel, following the substitution $r \rightarrow cr$. (a) Frequency-spatial distribution of LDOS. (b) Spatial dependence of LDOS near the coherence peak around the sub-gap. (c) Spatial dependence of LDOS at frequencies much higher than the coherence peak.}
	\label{fig_final}
\end{figure}

Finally, by combining the electronic Hamiltonian for the Bogoliubov quasiparticles in \eqnref{Hcfinal} and the wavefunction of Bogoliubov quasiparticles given in \eqnref{uvcons}, the total density of states \eqnref{rhototal} is given by:
\begin{eqnarray}\label{LDOSfinal}
\begin{aligned}
    \rho(\omega, r)=& \frac{1}{\pi} \frac{|\omega|}{\sqrt{-2 F^2+\omega^2}} \sum_{\sigma= \pm}\left[\frac{4 \omega_\sigma}{\pi} \operatorname{Si}\left(2 \omega_\sigma r\right)+\frac{2}{\pi} \frac{\cos \left(2 \omega_\sigma r\right)}{r}\right]\\
    &\times \left[1+\operatorname{sgn}(\omega) \frac{\left(2 \sqrt{2} F+\sigma \omega_\sigma\right)}{\sqrt{2} \sqrt{2 F^2+\left(\omega_\sigma+\sigma \sqrt{2} F\right)^2}}(-1)^r\right]
    \end{aligned}
\end{eqnarray}
as $\omega^2 > 2F^2$, while $\rho(\omega, r) = 0$ as $\omega^2 < 2F^2$. Here, $\omega_\sigma = -\sigma \sqrt{2} F + \sqrt{-2 F^2 + \omega^2}$ and $\operatorname{Si}(x) \equiv \int_0^x \frac{\sin t}{t} \, dt = \pi \sum_{l \geq 0} J_{l+1/2}^2(x/2)$. This hard gap of $2F$ in this spectrum is an artifact of the assumed exact nesting of the the CDW wavevector with the separation between the nodal points.
Note that in our simplified model, the induced charge order interaction only exists around the vortex and completely vanishes far away from the vortex and deep in the homogeneous SC part. In the latter case, the local density of states $\rho_{\infty}(\omega, r)$ can be obtained by setting $F = 0$ in \eqnref{LDOSfinal}, resulting in:
\begin{equation}\label{LDOSinf}
    \rho_{\infty}(\omega, r) \equiv \rho(\omega, r)_{F \rightarrow 0}= \frac{4}{\pi^2}\left[2 \omega \operatorname{Si}(2 \omega r)+\frac{\cos (2 \omega r)}{r}\right]
\end{equation}
which is consistent with the result obtained in Ref.~\cite{Bartosch.Nikolic.2006, Sachdev.Nikolic.2006, Mel2001}. However, note that the crossover regimes between them depend highly on microscopic details, which is not the focus of this work.
\begin{figure}[t]
	\centering
	\includegraphics[width=0.7\linewidth]{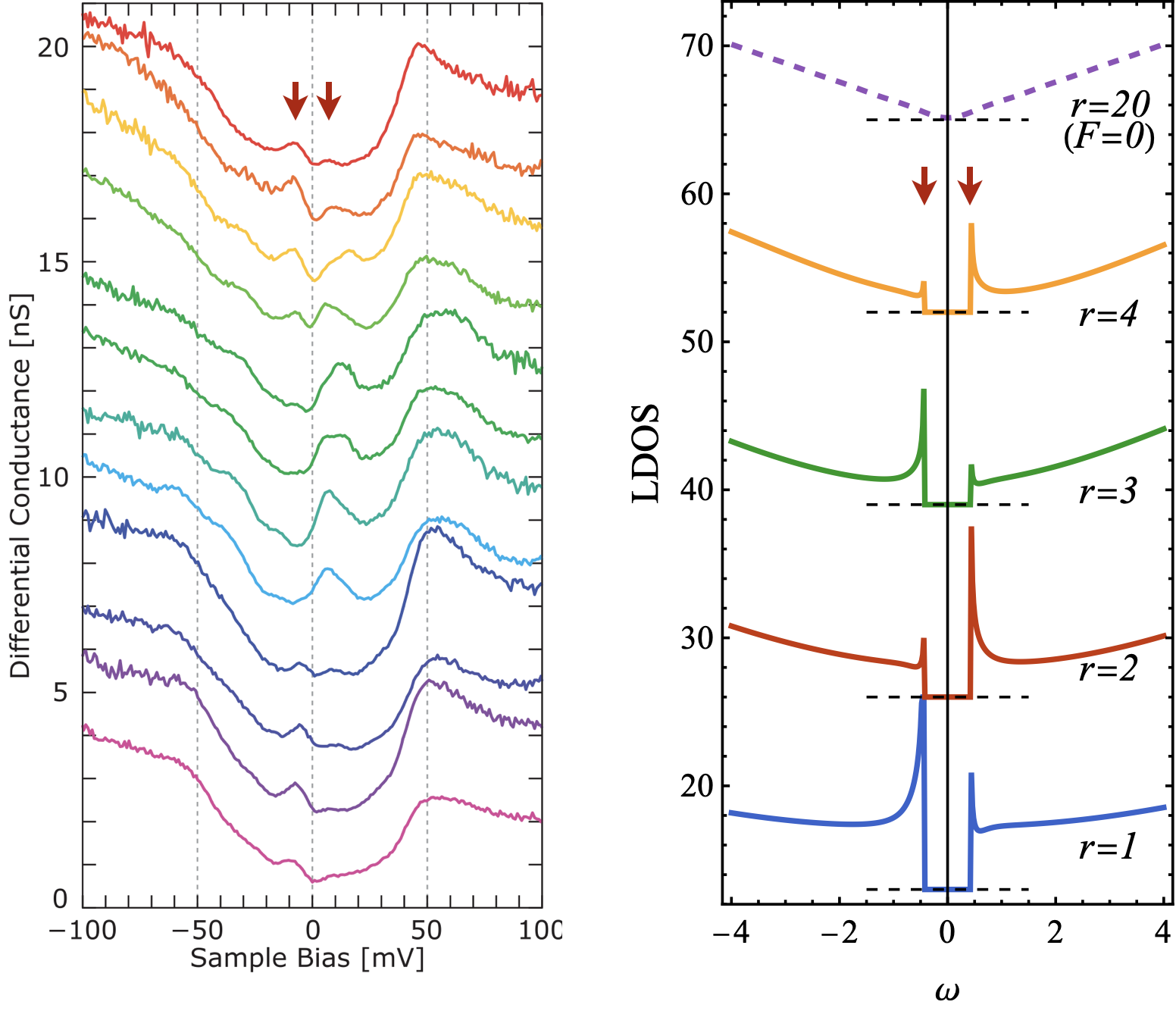}
	\caption{Left panel: experimental data from STM as reported in Fig.~2(c) of Ref. \cite{Nishida.Matsuba.2007} (red arrows added). Right panel: The total local density of states $\rho(r,\omega)$ in \eqnref{LDOSfinal} and \eqnref{LDOSinf}, following the substitution $r\rightarrow cr$. The observed sub-peaks in the STM are indicated with red arrows in both panels.}
	\label{fig_exp}
\end{figure}

Additionally, the inverse coordinate transformation in \eqnref{rotate} and \eqnref{scale} is also needed to fit the actual situation, resulting in $r \rightarrow cr$ along the $x$ direction, where $c \equiv \sqrt{v_F^{-2} / 2 + v_{\Delta}^{-2} / 2}$, which will be determined phenomenologically in realistic cases.

As a result, the total LDOS spectra after the replacement $r \rightarrow cr$ are shown in \figref{fig_final}, with the left panels corresponding to the physics near the vortex core, as described by \eqnref{LDOSfinal}, and the right panels corresponding to the physics far away from the vortex core, as described by \eqnref{LDOSinf}. From \figref{fig_final}(a), instead of a zero-bias peak, a gap of magnitude $\sqrt{2}F$ is observed around the vortex at zero frequency, accompanied by a coherent peak at the $\pm\sqrt{2}F$ frequency. This phenomenon originates from the interaction of the CDW potential coupling distinct gapless Dirac points, forming a gap. Note that this is the effective theory of Dirac points, so the observed sub-gap is actually inside the $d$-wave SC gap, meaning the SC Bogoliubov peak with higher energy is not shown here. Moreover, compared with the LDOS close to the coherent peak $\pm\sqrt{2}F$ [as shown in \figref{fig_final}(b)] of the sub-gap and far away from such peak [as shown in \figref{fig_final}(c)], one can find that, in proximity to the vortex core, the sub-gap spectrum exhibits apparent spatial modulation with a period of $2a$. More importantly, this modulation exhibits an `anti-phase' behavior for frequencies of opposite signs: this arises from the fact that the charge order induced amplitude for an electron near one Dirac point to convert to a hole near a different Dirac point, given in \eqnref{Ggamma} exhibits opposite signs for opposite frequencies, as discussed in Appendix \ref{App1}.
Furthermore, far from the vortex core, the density of states exhibits a featureless linear-$\omega$ dependency, which is shown in the dashed line of \figref{fig_exp}(b), and can be directly obtained via the expansion of \eqnref{LDOSinf} at the large $r$ limit. Such a result aligns with the characteristic `V'-shape behavior of $d$-wave superconductors.

Finally, the total local density of states $\rho(r,\omega)$ in \eqnref{LDOSfinal} and \eqnref{LDOSinf} can be compared with the vortex halo observed by STM \cite{Tamegai.Machida.2016, Nishida.Matsuba.2007, Nishida.Yoshizawa.2013, science.1066974} directly, as shown in \figref{fig_exp}.

\section{Discussion}

Our paper has considered the nature of vortices in a $d$-wave superconductor obtained by a confinement transition from the pseudogap metal. In terms of broken symmetry, and low energy quasiparticle spectrum, such a superconductor is identical to a $d$-wave superconductor obtained from a BCS pairing transition of a Fermi liquid with a Luttinger volume Fermi surface \cite{Christos24}. But, as we have shown here, the nature of the pseudogap does reveal itself in the short-distance physics in and around the vortex core.

As is well known, the conventional transition from the Fermi liquid to the $d$-wave superconductor is described by the Landau-Ginzburg theory of the condensation of a complex scalar field carrying charge $2e$, representing the superconducting order parameter. We have considered here a continuum theory of the pseudogap \cite{pnas.2302,CSSL24} in which the transition to the $d$-wave superconductor is described by the condensation of Higgs fields, the 4 charge $e$ complex scalars $B_{as}$ ($s=\pm$ is a valley index, and $a=1,2$ is a fundamental SU(2) gauge index), coupled to a SU(2) gauge field. Gauge-invariant bilinears of these Higgs fields realize the $d$-wave superconducting and period-2 bond density wave order parameters, as shown in Eq.~(\ref{eq:order_con}), and so we can loosely state that the Higgs fields are the `square roots' of the superconducting, charge density wave, and pair density wave order parameters. STM experiments observe 
LDOS modulations with periods 4 and 8 \cite{science.1066974,Fischer05,Nishida.Matsuba.2007,Nishida.Yoshizawa.2013,Tamegai.Machida.2016,Hamidian19,Hirschfeld20,Fujita20,Seamus21}. Our theory can be extended to these longer periods \cite{pnas.2302,BCS}, either by working with an extended lattice action for the $B_{a \vi}$, or by taking the continuum limit with additional valleys and continuum fields. We have left this computationally demanding task for the future, and worked here with the simpler situation with period 2 density waves for which significant analytic progress has been possible.   
We also mention earlier work on the vortex structure 
with 2 complex Higgs fields and a U(1) gauge theory, which did not have charge density wave 
order parameters \cite{SSvortex,NagaosaLee92,LeeWen01,LeeWenRMP}.

Our analysis should be contrasted with numerous earlier studies \cite{Ting01,Demler02,Ogata03,Ting03,BBBSS1,Schmid_2010,Agterberg15,Kivelson18,Pepin20,Hirschfeld20,TKLee23,Ghosal02,Ghosal24,Datta2023} in which the vortex core is described directly in terms of multiple SU(2) gauge-invariant order parameters, including superconductivity, pair and charge density waves. In such approaches, the normal state above $T_c$ is presumed to be amenable to be a description in a theory of these fluctuating orders, but in the explicit computations the state without any order parameters is a large Fermi surface Fermi liquid. We have instead assumed that the normal state is better described as a FL* metal with a background spin liquid: see Ref.~\cite{Mascot22} for a description of photoemission and STM in the normal state by such a theory. No symmetry is broken in the FL* state, and there is no direct reference to any particular fluctuating order. The multiple order parameters only make an explicit appearance when we consider the confinement of the spin liquid by the condensation of the Higgs field $B$.

Our main results appear in Figs.~\ref{fig_trivial}(d), \ref{fig_final}, and \ref{fig_exp}(b), along with experimental data in Fig.~\ref{fig_exp}(a). 
We considered a Higgs potential for $B$ so that the bulk homogeneous ground state is a $d$-wave superconductor. Nevertheless, as shown in 
Fig.~\ref{fig_trivial}(d), charge order emerges in the vortex core: this is a natural consequence of the associated suppression of superconductivity in the core, and the ability of the Higgs fields to rotate into the charge order direction (a complementary approach to this physics is provided by dual theories of quantum fluctuations of the vortices \cite{BBBSS05,BBS06,BS07}). Figs.~\ref{fig_final}(a) and \ref{fig_exp}(b) show the absence of the Wang-Macdonald zero-bias peak, and the associated appearance of peaks at non-zero bias in the vortex core. Finally, and most remarkably, these peaks have spatial modulations which are out-of-phase between positive and negative bias, as seen in Figs.~\ref{fig_final}(b) and \ref{fig_exp}(b). This is due to the Dirac structure of the quasiparticles in the $d$-wave superconductor, and matches the observations of Matsuba {\it et al.\/} \cite{Nishida.Matsuba.2007}, which we show in Fig.~\ref{fig_exp}(a).

Looking ahead, it is clearly important to extend this work to spatial modulations with longer periods, and understand the connection to models used for quantum oscillations \cite{BCS}. As we have noted earlier, the spatial modulations of the Higgs fields appear both in $\Delta_{\vi\vj}$ and $\rho_\vi$ (see Eq.~(\ref{eq:order_lattice})), and this can help understand the relative roles of charge and pair density wave order. The direct connection between the structure of the pseudogap metal and the vortex is a distinguishing feature of our theory, when compared to various competing order computations (whose normal state is a large Fermi surface Fermi liquid): this can be exploited to make a more quantitative connection between observations in the pseudogap and the vortex state.
Finally, a significant experimental and theoretical challenge is to understand the evolution of the vortex structure from low to high doping, as one proceeds from the parent pseudogap metal to the parent Fermi liquid \cite{Ancilla20,Ancilla20b}.

We also recall recent predictions of the spin liquid model of the pseudogap metal for other experimental probes. Ref.~\cite{Christos24} pointed out that the $d$-wave superconductor in the electron-doped cuprates should have nodal quasiparticles even though the normal state can be gapped along the Brillouin zone diagonals, and obtained a non-monotonic angular dispersion of the quasiparticles. Ref.~\cite{BCS} noted implications for RIXS experiments, which can detect spinons described by the critical $\mathbb{CP}^1$ theory.

\subsection*{Acknowledgements}

We thank Pietro Bonetti, Maine Christos, Seamus Davis, Peter Hirschfeld and Steve Kivelson for valuable discussions. This research was supported by the U.S. National Science Foundation grant No. DMR-2245246 and by the Simons Collaboration on Ultra-Quantum Matter which is a grant from the Simons Foundation (651440, S.S.). J.-X. Z. acknowledges the support from NSFC (Grant No. 12347107) and the Tsinghua Visiting Doctoral Students Foundation. 

\appendix
\section{Self-consistent calculation of $B$ field}\label{App0}

In this section, we will provide more details about the determination of the spatial modulation function in the ansatz for the chargon $B$ at the continuum limit, as shown in Sec. \ref{sec:continuum}.

\subsection{Flat configuration}\label{App01}
\begin{figure}[t]
	\centering
	\includegraphics[width=0.17\linewidth]{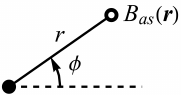}
	\caption{The schematic diagram of $r$ and $\phi$ in \eqnref{anaB}.}
	\label{fig_angle}
\end{figure}

With the flat configuration of $\vec{a}^z$, the ansatz for the chargon $B$ is given by \eqnref{anaB}. The free energy, derived by plugging this ansatz into \eqnref{totfree}, is given by: 
\begin{eqnarray}
    F&=&\int_0^{\infty} d r 2 r b^2\left[\frac{\partial f(r)}{\partial r}\right]^2+b^2 \frac{f^2(r)}{r}\left[1+\frac{\Phi_A(r)}{2 \pi}\right]^2+b^2 \frac{f^2(r)}{r}\left[1+\frac{\Phi_A(r)}{2 \pi}\right]^2\notag\\
    &\;&+2 m r b^2 f^2(r)+(4 u+v_3) r b^4 f^4(r)+\frac{1}{g} \frac{1}{r}\left(\frac{\partial}{\partial_r} \Phi_A(r)\right)^2\label{Fr}
\end{eqnarray}
where $f(r) \in [0,1]$ and $\phi$ is the angle in real space (as illustrated in \figref{fig_angle}), encircling the pole. The equations of motion derived from \eqnref{Fr} are:
\begin{eqnarray}\label{EOF1}
&\;&-2 \frac{\partial f(r)}{\partial r}-2 r \frac{\partial^2 f(r)}{\partial r^2}+\frac{f(r)}{r}\left[1+\frac{\Phi_A(r)}{2 \pi}\right]^2+\frac{f(r)}{r}\left[1+\frac{\Phi_A(r)}{2 \pi}\right]^2\notag\\
&\;&\;\;\;\;\;\;\;\;\;\;\;\;\;\;\;\;\;\;\;\;\;\;\;\;\;\;\;\;\;\;\;\;\;\;\;\;\;\;\;\;+2 m r f(r)+2(4 u+v_3) r b^2 f^3(r)=0
\end{eqnarray}
\begin{eqnarray}
    2 b^2 \frac{r f^2(r)}{2 \pi}\left[n_{+}+\frac{\Phi_A(r)}{2 \pi}\right]+2 b^2 \frac{r f^2(r)}{2 \pi}\left[n_{-}+\frac{\Phi_A(r)}{2 \pi}\right]+\frac{2}{g} \frac{\partial}{\partial r} \Phi(r)-\frac{2 r}{g} \frac{\partial^2 \Phi(r)}{\partial r^2}=0\label{EOF2}
\end{eqnarray}
with the boundary conditions:
\begin{eqnarray}
\begin{aligned}\label{BC1}
&f(r) \xrightarrow{r \rightarrow 0} 0,\;\;\;\;\;\;\;f(r) \xrightarrow{r \rightarrow \infty} 1 \\
&\Phi_A(r) \xrightarrow{r \rightarrow 0} 0,\;\;\;\;\; \Phi_A(r) \xrightarrow{r \rightarrow \infty} 2 \pi.
\end{aligned}
\end{eqnarray}
The numerical results derived from \eqnref{EOF1} and \eqnref{EOF2} under the boundary conditions \eqnref{BC1} are depicted in \figref{fig_trivial}(a).

\subsection{Vison-like configuration}\label{App02}
Similarly, with the vison-like configuration of $\vec{a}^z$, the ansatz for the chargon $B$ is given by \eqnref{Bpm2}. The free energy, derived by plugging this ansatz into \eqnref{totfree}, is given by:
\begin{eqnarray}
    F&=&\int_0^{\infty} d rb^2 r\left[\frac{\partial f(r)}{\partial r}\right]^2+b^2 r\left[\frac{\partial h(r)}{\partial r}\right]^2+b^2 r \frac{f^2(r)}{r^2}\left[-1+\frac{\Phi_A(r)}{2 \pi}+\frac{\Phi_a(r)}{2 \pi}\right]^2+b^2 r \frac{h^2(r)}{r^2}\left[\frac{\Phi_A(r)}{2 \pi}-\frac{\Phi_a(r)}{2 \pi}\right]^2\notag\\
    &\;&+b^2 r m\left(f^2(r)+h^2(r)\right)+b^4 r u\left(f^2(r)+h^2(r)\right)^2+v_1 b^4 r\left[f^2(r)-h^2(r)\right]^2+v_3 b^4 r f^2(r) h^2(r)\frac{\partial f(r)}{\partial r}\\
    &\;&+\frac{1}{g} \frac{1}{r}\left(\frac{\partial \Phi_A(r)}{\partial_r} \right)^2+\frac{1}{\lambda} \frac{1}{r}\left(\frac{\partial \Phi_a(r)}{\partial_r} \right)^2 \notag
\end{eqnarray}
The corresponding equations of motion are:
\begin{eqnarray}\label{EQ1}
&\;&-\frac{1}{r} \frac{\partial f(r)}{\partial r}-\frac{\partial^2 f(r)}{\partial r^2}+\frac{f(r)}{r^2}\left[-1+\frac{\Phi_A(r)}{2 \pi}+\frac{\Phi_a(r)}{2 \pi}\right]^2+m f(r)+2 b^2 u\left(f^2(r)+h^2(r)\right) f(r) \notag
\\ &\;&\;\;\;\;\;\;\;\;\;\;\;\;\;\;\;\;\;\;\;\;\;\;\;\;\;\;\;\;\;\;\;\;\;\;\;\;\;\;\;\;\;\;\;\;\;\;\;\;\;\;\;\;\;\;\;\;\;\;\;\;\;\;\;\;\;\;\;\;\;\;+2 v_1 b^2\left[f^2(r)-h^2(r)\right] f(r)+v_3 b^2 f(r) h^2(r)=0
\end{eqnarray}

\begin{eqnarray}
    &\;&-\frac{1}{r} \frac{\partial h(r)}{\partial r}-\frac{\partial^2 h(r)}{\partial r^2}+\frac{h(r)}{r^2}\left[\frac{\Phi_A(r)}{2 \pi}-\frac{\Phi_a(r)}{2 \pi}\right]^2+m h(r)+2 b^2 u\left(f^2(r)+h^2(r)\right) h(r) \notag\\
    &\;&\;\;\;\;\;\;\;\;\;\;\;\;\;\;\;\;\;\;\;\;\;\;\;\;\;\;\;\;\;\;\;\;\;\;\;\;\;\;\;\;\;\;\;\;\;\;\;\;\;\;\;\;\;\;\;\;\;\;\;\;\;\;\;\;\;\;\;\;\;\;-2 v_1 b^2\left[f^2(r)-h^2(r)\right] h(r)+v_3 b^2 h(r) f^2(r)=0
\end{eqnarray}

\begin{eqnarray}
        &\;&\frac{b^2 r}{2 \pi} f^2(r)\left[-1+\frac{\Phi_A(r)}{2 \pi}+\frac{\Phi_a(r)}{2 \pi}\right]-\frac{b^2 r}{2 \pi} h^2(r)\left[\frac{\Phi_A(r)}{2 \pi}-\frac{\Phi_a(r)}{2 \pi}\right]+\frac{1}{\lambda} \frac{\partial \Phi_a(r)}{\partial r}-\frac{r}{\lambda} \frac{\partial^2 \Phi_a(r)}{\partial r^2}=0
\end{eqnarray}    

\begin{eqnarray}\label{EQ4}
    &\;&\frac{b^2 r}{2 \pi} f^2(r)\left[-1+\frac{\Phi_A(r)}{2 \pi}+\frac{\Phi_a(r)}{2 \pi}\right]+\frac{b^2 r}{2 \pi} h^2(r)\left[\frac{\Phi_A(r)}{2 \pi}+\frac{\Phi_a(r)}{2 \pi}\right]+\frac{1}{g} \frac{\partial}{\partial r} \Phi_A(r)-\frac{r}{g} \frac{\partial^2 \Phi_A(r)}{\partial r^2}=0
    \end{eqnarray}
with the boundary conditions:
\begin{eqnarray}\label{BCC1}
\begin{aligned}
    &f(r) \xrightarrow{r \rightarrow 0} 0, \;\;\;\;\;\;f(r) \xrightarrow{r \rightarrow \infty} 1, \\
    &h(r) \xrightarrow{r \rightarrow 0} h_0,\;\;\;\;\; h(r) \xrightarrow{r \rightarrow \infty} 1\\ &\Phi_A(r) \xrightarrow{r \rightarrow 0} 0,\;\;\;\; \Phi_A(r) \xrightarrow{r \rightarrow \infty}  \pi,\\
    &\Phi_a(r) \xrightarrow{r \rightarrow 0} 0, \;\;\;\;\;\Phi_a(r) \xrightarrow{r \rightarrow \infty} \pi.
\end{aligned}
    \end{eqnarray}
The value $b$ can be deduced from \eqnref{EQ1} in the large $r$ limit, resulting in $b^2 = \frac{-m}{(4u+v)}$. Also, $h_0$ is set such that $\partial_r h(r) \xrightarrow{r \rightarrow 0} 0$. The numerical results derived from Eqs.~(\ref{EQ1})-(\ref{EQ4}) under the boundary conditions \eqnref{BCC1} are shown in \figref{fig_trivial}(c).

\section{Derivation of LDOS Near the Vortex}\label{App1}
In this section, we will provide more details about the determination of LDOS $\rho(\omega, r)$ in \eqnref{LDOSfinal}.

\begin{figure}[t]
	\centering
	\includegraphics[width=0.45\linewidth]{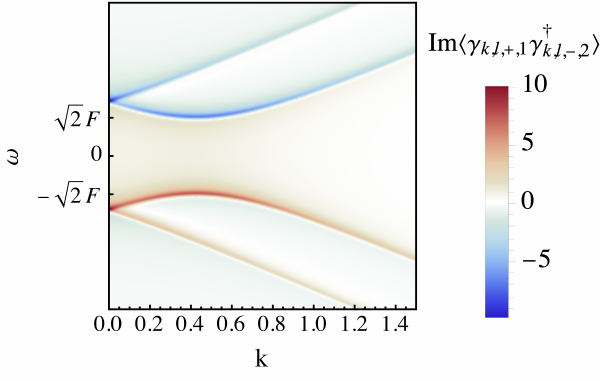}
	\caption{With parameters $F=0.3$ and $\Gamma=0.03$, the imaginary part of $\left\langle\gamma_{k, l,+, 1} \gamma_{k, l,-,2}^{\dagger}\right\rangle$ for $l>0$ from \eqnref{Ggamma} after the replacement of $\omega \rightarrow \omega + i\Gamma$.
}
	\label{fig_gammaF}
\end{figure}

Firstly, based on the electronic Hamiltonian \eqnref{Hcfinal}, the Green's function of Bogoliubov quasiparticles $\gamma$ is given by:
\begin{eqnarray}\label{Ggamma}
    \left\langle\Psi_{k, l}^{(\alpha)} \otimes\left(\Psi_{k, l}^{(\alpha)}\right)^{\dagger}\right\rangle=-\left(\omega-V_{k, l}^{(\alpha)}\right)^{-1}=\frac{M^{(\alpha)}}{16 F^4-8 F^2 \omega^2+\left(k^2-\omega^2\right)^2}
\end{eqnarray}
where $\alpha = 1, 3$ and $M^{(\alpha)}$ represents a $4 \times 4$ matrix with the expression:
\begin{eqnarray}
    M^{(\alpha)}=4 F^2\left(\omega \sigma_{00}+k \sigma_{10}\right)&+&\left(k^2-\omega^2\right)\left(\omega \sigma_{00}+k \sigma_{30}\right)-(\alpha -2)F s_l\left(4 F^2-\omega^2\right)\left(-\sigma_{11}-\sigma_{12}+\sigma_{31}+\sigma_{32}\right)\notag\\
    &+&(\alpha -2)F s_l\left(2 k \omega\left(\sigma_{01}+\sigma_{02}\right)+k^2\left(\sigma_{11}+\sigma_{12}+\sigma_{31}+\sigma_{32}\right)\right)
\end{eqnarray}
Here, we present the imaginary part of $\left\langle \gamma_{k,l, +, 1} \gamma_{k,l, -, 2}^{\dagger} \right\rangle$, which corresponds to the amplitude for an electron near the $\alpha=1$ Dirac point converting to a hole near the $\alpha=2$ Dirac point, in \figref{fig_gammaF} as an example. The sign of the amplitude is opposite for opposite frequencies, and all the weight is inside the gap $-\sqrt{2}F < \omega < \sqrt{2}F$.

Starting from \eqnref{rhototal}, the LDOS $\rho(\omega, r)$ is contributed by both the uniform part $\rho_{\text{uni}}(\omega, r)$ and the modulation part $\rho_{\text{CDW}}(\omega, r)$. In the following, we will discuss these two parts of the contribution separately.

Using the wavefunction of Bogoliubov quasiparticles given in \eqnref{uvcons}, the uniform part $\rho_{\text{uni}}(\omega, r)$ can be further expressed as:
\begin{eqnarray}
    \rho_{\mathrm{uni}}(\omega, r) &=& \frac{2}{\pi} \sum_{\alpha=1}^4 \operatorname{Im}\left\langle c_{\alpha, \uparrow}(r) c_{\alpha, \uparrow}^{\dagger}(r)\right\rangle \notag\\
    &=& \frac{2}{\pi}\sum_{\alpha=1}^4 \int_0^{\infty} d k \sum_l \sum_{q, q^{\prime}= \pm}\left[u_{k, l, q, \alpha}(r)-i v_{k, l, q, \alpha}(r)\right]\left[u_{k, l, q^{\prime}, \alpha}^*(r)+i v_{k, l, q^{\prime}, \alpha}^*(r)\right] \Im \left\langle\gamma_{k, l, q, \alpha} \gamma_{k, l, q^{\prime}, \alpha}^{\dagger}\right\rangle \notag\\
    &=& \frac{2}{\pi^2}\int_0^{\infty} d k \sum_l\left[4 \sum_{l \geq 0} J_{l+1 / 2}^2(k r)+\frac{2}{\pi} \frac{\cos (2 k r)}{k r}\right]\sum_{\sigma_1, \sigma_2= \pm} \Im  \frac{-k}{2\left(\omega+i0^+-\sigma_1 \sqrt{2 F^2+\left(k+\sigma_2 \sqrt{2} F\right)^2}\right)} \notag\\
    &=& \frac{2}{\pi}\int_0^{\infty} d k \left[\frac{2 k}{\pi} \operatorname{Si}(2 k r)+\frac{1}{\pi} \frac{\cos (2 k r)}{r}\right] \sum_{\sigma_1, \sigma_2= \pm} \delta\left(\omega-\sigma_1 \sqrt{2 F^2+\left(k+\sigma_2 \sqrt{2} F\right)^2}\right)\label{rhouni}
\end{eqnarray} 
where $\left\langle \gamma_{k, l, q, \alpha} \gamma_{k, l, q^{\prime}, \alpha}^{\dagger} \right\rangle$ can be obtained from \eqnref{Ggamma}.

\begin{figure}[t]
	\centering
	\includegraphics[width=1\linewidth]{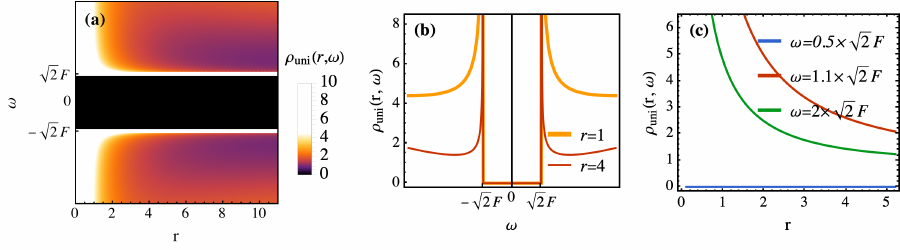}
	\caption{The uniform component of the local density of states $\rho_{\text{uni}}(r,\omega)$ in \eqnref{rhounifinal}, following the substitution $r\rightarrow cr$. This is considered with the parameters $c=1/10$ and $F=0.3$.}
	\label{fig_rhouni}
\end{figure}

According to the relation $\delta[g(x)] = \sum_i \frac{\delta \left( x - x_i \right)}{\left| g^{\prime} \left( x_i \right) \right|}$, where $x_i$ denotes the roots of $g$, the following equation can be derived:
\begin{equation}\label{delta1}
    \delta\left(\omega\pm\sqrt{2 F^2+(k+\sigma \sqrt{2} F)^2}\right)=\left|\frac{\omega}{\sqrt{2 F^2+\omega^2}}\right| \sum_{\sigma^{\prime}=\pm} \delta\left(k-\left(-\sigma \sqrt{2} F+\sigma^{\prime} \sqrt{-2 F^2+\omega^2}\right)\right).
\end{equation}
By referring to Table \ref{tab:sign}, which illustrates the signs of $\pm \sqrt{2} F \pm \sqrt{-2 F^2 + \omega^2}$, we can analyze the values of $\rho_{\text{uni}}(\omega, r)$ within different regions of $\omega$. In the case of $\omega > \sqrt{2} F$,

\begin{table}
    \centering
    \begin{tabular}{c|c|c}
         & $\omega>\sqrt{2}F$ & $\omega<\sqrt{2}F$\\
         \hline
        $\sqrt{2} F+\sqrt{\omega^2-2 F^2}$ & $+$ & $-$\\
        \hline
        $\sqrt{2} F-\sqrt{\omega^2-2 F^2}$ & $-$ & $-$ \\
        \hline
        $-\sqrt{2} F+\sqrt{\omega^2-2 F^2}$ & $+$ & $+$ \\
        \hline
        $-\sqrt{2} F-\sqrt{\omega^2-2 F^2}$ & $-$ & $-$ \\
    \end{tabular}
    \caption{The sign of $\pm \sqrt{2} F \pm \sqrt{-2 F^2+\omega^2}$ in different regions of $\omega$ }
    \label{tab:sign}
\end{table}

\begin{equation}
    \sum_{\sigma_1, \sigma_2= \pm} \delta\left(\omega-\sigma_1 \sqrt{2 F^2+\left(k+\sigma_2 \sqrt{2} F\right)^2}\right)=\sum_{\sigma= \pm}\left|\frac{\omega}{\sqrt{-2 F^2+\omega^2}}\right| \delta\left(k-\left(-\sigma \sqrt{2} F+\sqrt{-2 F^2+\omega^2}\right)\right)
\end{equation}
Consequently, the expression for $\rho_{\text{uni}}(\omega, r)$ in \eqnref{rhouni} is:
\begin{eqnarray}
    \rho_{\text{uni}}(\omega,r)=\frac{1}{\pi}\left|\frac{\omega}{\sqrt{-2 F^2+\omega^2}}\right| \sum_{\sigma= \pm}\left[\frac{4 \omega_\sigma}{\pi} \operatorname{Si}\left(2 \omega_\sigma r\right)+\frac{2}{\pi} \frac{\cos \left(2 \omega_\sigma r\right)}{r}\right]
\end{eqnarray}
where
\begin{equation}
    \omega_\sigma \equiv-\sigma \sqrt{2} F+\sqrt{\omega^2-2 F^2}.
\end{equation}
In the scenario where $\omega < \sqrt{2} F$, a similar approach leads to the conclusion that $\rho_{\text{uni}}(\omega, r)$ remains as defined in the above equation. However, for $-\sqrt{2} F < \omega < \sqrt{2} F$, $\omega - \sigma_1 \sqrt{2 F^2 + \left( k + \sigma_2 \sqrt{2} F \right)^2} \neq 0$ consistently, implying that:
\begin{equation}
    \sum_{\sigma_1, \sigma_2= \pm} \delta\left(\omega-\sigma_1 \sqrt{2 F^2+\left(k+\sigma_2 \sqrt{2} F\right)^2}\right)=0,
\end{equation}
and therefore,
\begin{equation}
    \rho_{\text {uni }}(\omega, r)=0
\end{equation}
As a result, $\rho_{\text{uni}}(\omega, r)$ can be formulated as:
\begin{equation}
    \rho_{\text {uni }}(\omega, r)= \begin{cases}\frac{1}{\pi}\left|\frac{\omega}{\sqrt{-2 F^2+\omega^2}}\right| \sum_{\sigma= \pm}\left[\frac{4 \omega_\sigma}{\pi} \operatorname{Si}\left(2 \omega_\sigma r\right)+\frac{2}{\pi} \frac{\cos \left(2 \omega_\sigma r\right)}{r}\right] & \omega^2>2 F^2 \\ 0 & \omega \leq 2 F^2\end{cases}.\label{rhounifinal}
\end{equation}
The results of \eqnref{rhounifinal} after substituting $r \rightarrow cr$, are depicted in \figref{fig_rhouni}. This illustration suggests the opening of a gap with a magnitude of $\sqrt{2} F$, leading to a coherent peak at the frequency $\sqrt{2} F$. Notably, LDOS exhibits a tendency towards divergence as it approaches the vortex core.

On the other hand, the modulation part $\rho_{\text{CDW}}(\omega, r)$ can be further given by:
\begin{eqnarray}
    \rho_{\mathrm{CDW}}(\omega, r) &=& \frac{2}{\pi}  \operatorname{Im}\left[\left\langle c_{1, \sigma}(r) c_{2, \sigma}^{\dagger}(r)\right\rangle+1 \leftrightarrow 2+\left\langle c_{3, \sigma}(r) c_{4, \sigma}^{\dagger}(r)\right\rangle+3 \leftrightarrow 4\right]\notag\\
    &=&\frac{2}{\pi^2} \int_0^{\infty} d k\left[\frac{4}{\pi} \operatorname{Si}(2 k r)+\frac{2}{\pi} \frac{\cos (2 k r)}{k r}\right] \operatorname{Im} \frac{4 F k\left(4 F^2-v^2\right)}{\prod_{\sigma_1, \sigma_2= \pm}\left(\omega+i0^+-\sigma_1 \sqrt{2 F^2+\left(k+\sigma_2 \sqrt{2} F\right)^2}\right)}\notag\\
    &=&-\frac{1}{\pi} \int_0^{\infty} d k\left[\frac{4 k}{\pi} \operatorname{Si}(2 k r)+\frac{2}{\pi} \frac{\cos (2 k r)}{r}\right] \sum_{\sigma_1, \sigma_2= \pm} \frac{\left(2 \sqrt{2} F+\sigma_1 k\right)}{\sqrt{2} \sqrt{2 F^2+\left(k+\sigma_1 \sqrt{2} F\right)^2}} \sigma_2 \notag\\
    &\;&\times\delta\left(\omega+\sigma_2 \sqrt{2 F^2+\left(k+\sigma_1 \sqrt{2} F\right)^2}\right)
\end{eqnarray}
From \eqnref{delta1} and Table \ref{tab:sign}, we can discuss the value of $\rho_{\text{CDW}}(\omega, r)$ at different $\omega$ regions. Specifically, when $\omega > \sqrt{2} F$,
\begin{eqnarray}
    &\;&\sum_{\sigma_1, \sigma_2= \pm} \frac{\left(2 \sqrt{2} F+\sigma_1 k\right)}{\sqrt{2} \sqrt{2 F^2+\left(k+\sigma_1 \sqrt{2} F\right)^2}} \sigma_2
    \delta\left(\omega+\sigma_2 \sqrt{2 F^2+\left(k+\sigma_1 \sqrt{2} F\right)^2}\right)\notag\\
    &=&-\sum_{\sigma_1= \pm} \frac{\left(2 \sqrt{2} F+\sigma_1 k\right)}{\sqrt{2} \sqrt{2 F^2+\left(k+\sigma_1 \sqrt{2} F\right)^2}}
    \times\left|\frac{\omega}{\sqrt{-2 F^2+\omega^2}}\right| \delta\left(k-\left(-\sigma_1 \sqrt{2} F+\sqrt{-2 F^2+\omega^2}\right)\right)
\end{eqnarray}
thus,
\begin{equation}
    \rho_{\mathrm{CDW}}(\omega, r)=\frac{1}{\pi} \sum_{\sigma= \pm}\left[\frac{4 \omega_\sigma}{\pi} \operatorname{Si}\left(2 \omega_\sigma r\right)+\frac{2}{\pi} \frac{\cos \left(2 \omega_\sigma r\right)}{r}\right] \frac{\left(2 \sqrt{2} F+\sigma \omega_\sigma\right)}{\sqrt{2} \sqrt{2 F^2+\left(\omega_\sigma+\sigma \sqrt{2} F\right)^2}} \frac{|\omega|}{\sqrt{-2 F^2+\omega^2}}
\end{equation}
where
\begin{equation}
    \omega_\sigma =-\sigma \sqrt{2} F+\sqrt{-2 F^2+\omega^2}
\end{equation}

When $\omega <\sqrt{2}F$,
\begin{eqnarray}
    &\;&\sum_{\sigma_1, \sigma_2= \pm} \frac{\left(2 \sqrt{2} F+\sigma_1 k\right)}{\sqrt{2} \sqrt{2 F^2+\left(k+\sigma_1 \sqrt{2} F\right)^2}} \sigma_2
    \delta\left(\omega+\sigma_2 \sqrt{2 F^2+\left(k+\sigma_1 \sqrt{2} F\right)^2}\right)\notag\\
    &=&\sum_{\sigma_1
    = \pm} \frac{\left(2 \sqrt{2} F+\sigma_1 k\right)}{\sqrt{2} \sqrt{2 F^2+\left(k+\sigma_1 \sqrt{2} F\right)^2}}\times\left|\frac{\omega}{\sqrt{-2 F^2+\omega^2}}\right| \delta\left(k-\left(-\sigma_1 \sqrt{2} F+\sqrt{-2 F^2+\omega^2}\right)\right)
\end{eqnarray}
thus,
\begin{equation}
    \rho_{\mathrm{CDW}}(\omega, r)=-\frac{1}{\pi} \sum_{\sigma= \pm}\left[\frac{4 \omega_\sigma}{\pi} \operatorname{Si}\left(2 \omega_\sigma r\right)+\frac{2}{\pi} \frac{\cos \left(2 \omega_\sigma r\right)}{r}\right] \frac{\left(2 \sqrt{2} F+\sigma \omega_\sigma\right)}{\sqrt{2} \sqrt{2 F^2+\left(\omega_\sigma+\sigma \sqrt{2} F\right)^2}} \frac{|\omega|}{\sqrt{-2 F^2+\omega^2}}
\end{equation}

\begin{figure}[t]
	\centering
	\includegraphics[width=1\linewidth]{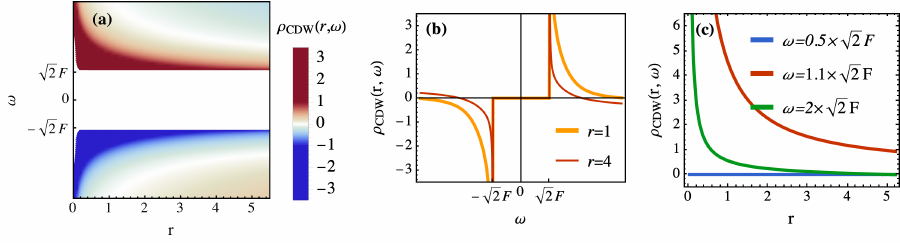}
	\caption{The charge modulation component of the local density of states $\rho_{\text{CDW}}(r,\omega)$ in \eqnref{rhoCDWfinal}, following the substitution $r\rightarrow cr$. This is considered with the parameters $c=1/10$ and $F=0.3$.}
	\label{fig_rhoCDW}
\end{figure}

When $-\sqrt{2} F < \omega < \sqrt{2} F$, $\omega - \sigma_1 \sqrt{2 F^2 + \left( k + \sigma_2 \sqrt{2} F \right)^2} \neq 0$ all the time, so that
\begin{equation}
    \sum_{\sigma_1, \sigma_2= \pm} \delta\left(\omega-\sigma_1 \sqrt{2 F^2+\left(k+\sigma_2 \sqrt{2} F\right)^2}\right)=0,
\end{equation}
yielding
\begin{equation}
    \rho_{\text {CDW}}(\omega, r)=0
\end{equation}
As a result, the expression for $\rho_{\text{CDW}}(\omega, r)$ can be summarized as:
\begin{eqnarray}\label{rhoCDWfinal}
    \rho_{\mathrm{CDW}}(\omega, r)= \begin{cases}\frac{1}{\pi}\left|\frac{\omega}{\sqrt{-2 F^2+\omega^2}}\right| \sum_{\sigma= \pm}\left[\frac{4 \omega_\sigma}{\pi} \operatorname{Si}\left(2 \omega_\sigma r\right)+\frac{2}{\pi} \frac{\cos \left(2 \omega_\sigma r\right)}{r}\right] \frac{\operatorname{sgn}(\omega)\left(2 \sqrt{2} F+\sigma \omega_\sigma\right)}{\sqrt{2} \sqrt{2 F^2+\left(\omega_\sigma+\sigma \sqrt{2} F\right)^2}} & \text { as } \omega^2>2 F^2 \\ 0 & \text { as } \omega^2 \leq 2 F^2\end{cases}
\end{eqnarray}
where
\begin{equation}
    \omega_\sigma =-\sigma \sqrt{2} F+\sqrt{-2 F^2+\omega^2}
\end{equation}
The result of \eqnref{rhoCDWfinal} after the substitution $r \rightarrow cr$ is shown in \figref{fig_rhoCDW}, indicating that coherent peaks at the frequency $\sqrt{2} F$ exhibit opposite signs. Note that the formation of a peak is associated with the formation of a gap, while the opposite sign arises from the opposite amplitude for an electron converting to a hole at distinct Dirac points for opposite frequency, as illustrated in \figref{fig_gammaF}. Such opposite signs of amplitude, combined with the modulation factor $(-1)^r$ in \eqnref{rhototal}, lead to the out-of-phase modulation between positive and negative frequencies.

Combining the results of $\rho_{\text{uni}}$ in \eqnref{rhounifinal} and $\rho_{\text{CDW}}$ in \eqnref{rhoCDWfinal}, one obtains the LDOS $\rho$ in \eqnref{LDOSfinal}.

\begin{figure}[t]
	\centering
	\includegraphics[width=1\linewidth]{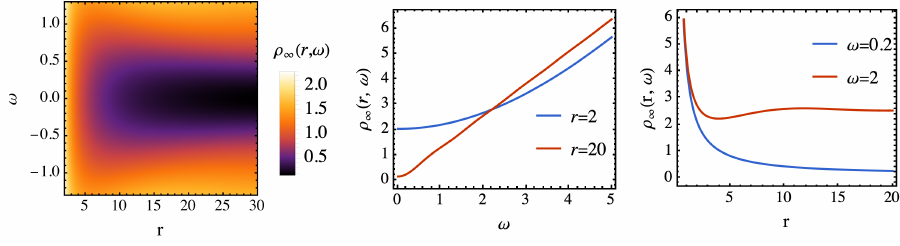}
	\caption{Local density of states without charge order potential, $\rho_{\text{LDOS}}(r,\omega)$, in \eqnref{rhono} with the substitution $r \rightarrow cr$. Parameters: $c=1/10$.}
	\label{LDOS}
\end{figure}

On the other hand, the LDOS without the charge order potential can be obtained by setting $F=0$, which leads to
 \begin{equation}\label{rhono}
    \rho_{\infty}(\omega, r) \equiv \rho(\omega, r)_{F \rightarrow 0}=\frac{4}{\pi^2}\left[2 \omega \operatorname{Si}(2 \omega r)+\frac{\cos (2 \omega r)}{r}\right],
\end{equation}
as in \eqnref{LDOSinf}. The result of \eqnref{rhono} is shown in \figref{LDOS}, indicating the enhancement of density as it approaches the vortex core ($r \rightarrow 0$), while there is no significant peak at zero bias. Importantly, far from the vortex core, the LDOS exhibits a linear dependence on frequency, which is characteristic of $d$-wave superconductors.

\bibliography{main3}
\end{document}